\documentclass[conference]{IEEEtran}
\IEEEoverridecommandlockouts
\usepackage[protrusion=true,expansion=true]{microtype}
\usepackage{cite}
\usepackage{amsmath,amssymb,amsfonts}
\usepackage{algorithmic}
\usepackage{graphicx}
\usepackage[hidelinks]{hyperref}
\usepackage{orcidlink}
\usepackage{bbm}
\usepackage{balance}
\usepackage{textcomp}
\usepackage{xcolor}
\def\BibTeX{{\rm B\kern-.05em{\sc i\kern-.025em b}\kern-.08em
    T\kern-.1667em\lower.7ex\hbox{E}\kern-.125emX}}

\usepackage{algorithm}

\usepackage{braket}
\usepackage{quantikz}

\usepackage{tikz}

\newcommand\copyrighttext{%
  \footnotesize \textcopyright 2026 IEEE. Personal use of this material is permitted.
  Permission from IEEE must be obtained for all other uses, in any current or future
  media, including reprinting/republishing this material for advertising or promotional
  purposes, creating new collective works, for resale or redistribution to servers or
  lists, or reuse of any copyrighted component of this work in other works.}
\newcommand\copyrightnotice{%
\begin{tikzpicture}[remember picture,overlay]
\node[anchor=south,yshift=10pt] at (current page.south) 
  {\fbox{\parbox{\dimexpr\textwidth-\fboxsep-\fboxrule\relax}{\copyrighttext}}};
\end{tikzpicture}%
}

\begin{document}

\title{High-level quantum structured programs \\ as quantum registers compositions
\thanks{This works has been partially funded by the EU (Next GenerationEU/PRTR), by the Spanish MICIU (PDC2025-165096-C31). It is also supported by project ATHENA (PID2024-155693NB-C41, MICIU/AEI/10.13039/501100011033); and 85\% co-funded by the EU, ERDF and the Regional Government of Extremadura. Managing Authority: Spanish Ministry of Finance, Aid GR24099 and Project IB24190}}

\author{\IEEEauthorblockN{David Chamizo\orcidlink{0009-0009-4659-5701}}
\IEEEauthorblockA{\textit{Quercus Software Engineering Group} \\
\textit{University of Extremadura}\\
Cáceres, Spain \\
davidchs@unex.es}
\and \IEEEauthorblockN{Jose Garcia-Alonso\orcidlink{0000-0002-6819-0299}}
\IEEEauthorblockA{\textit{Quercus Software Engineering Group} \\
\textit{University of Extremadura}\\
Cáceres, Spain \\
jgaralo@unex.es}
\and \IEEEauthorblockN{Juan M. Murillo\orcidlink{0000-0003-4961-4030}}
\IEEEauthorblockA{\textit{Quercus Software Engineering Group} \\
\textit{University of Extremadura}\\
Cáceres, Spain \\
juanmamu@unex.es}
}

\maketitle
\copyrightnotice

\vspace{-1em}\begin{abstract}
Current quantum programs are mainly designed at the level of quantum gates acting on individual qubits; on a large scale and for complex problems this may involve a high cognitive load on the programmer, making the program specification nontrivial and error-prone. In this context, providing quantum programming with higher abstraction mechanisms will assist in making this task more manageable and robust against design errors. In this work, a conceptual framework is addressed following the notion of the whole quantum computation as a structure composed of quantum registers representing each an undivided entity. Thus, computation progresses through semantically well-defined transformations that act on, or entangle, quantum registers, thereby modifying the global state. Ultimately, the program reaches the desired state by following a specific composition strategy. With this in mind, high-level syntax is presented through an algebraic formalism that bridges them with their low-level semantics. Proposed syntax is based on certain well-know operations used on quantum algorithms that apply phase shifts upon logical condition satisfaction or leverage on parallel evaluation. Based solely on the formalized operations, a quantum satisfiability modulo theories (SMT) solver can be designed. At its core, this work contributes to establishing some methodological principles towards realizing a high-level quantum structured programming.
\end{abstract}


\section{Introduction}
Quantum computing was conceived through the integration of quantum mechanics into computational models~\cite{feynman1982simulating,deutsch1985quantum} aiming to create quantum computer programs that enable a more efficient computation than any previously known classical computation~\cite{grover1996search,shor1999polynomial,arute2019quantum}. Currently, research is mainly focused on enabling its hardware~\cite{de2021materials,oukaira2025qhd}, designing novel quantum algorithms and error-correcting codes~\cite{dalzell2023quantum,brun_2020}, and establishing the foundations of quantum software engineering~\cite{murillo2025qse}. However, while quantum hardware continues to progress at a remarkable pace~\cite{abughanem2024ibm,jiang2026correction}, quantum software, and specifically quantum programming, is not advancing commensurately. As a consequence, the advent of practical quantum computing is hindered. 

As stated by Zhao~\cite{zhao2020quantum}, quantum programming can be understood as ``the process of designing and building executable quantum computer programs to achieve a particular computing result''. At present, building a quantum program is mostly restricted to programming languages that are devised for implementing quantum gates on individual qubits~\cite{heim2020quantum,abhari2012scaffold,green2013quipper,svore2018q,bergholm2022pennyLane,qiskit2024,cirq2025}. As a result, quantum program design is realized at a low level, comparable to classical assembler languages~\cite{wilkes1958preparation}, making an intricate computation a complex task and error-prone. In view of this, quantum programs ought to be adjusted to an abstraction level adapted to the dimension of the problems being addressed. This would not only benefit simplification but also verifiability, modularity, scalability and expressive capacity of quantum programs~\cite{di2025art}. At the same time, an appropriate abstraction level may prove advantageous in the design and implementation of novel quantum algorithms. Nevertheless, raising the level of abstraction in quantum programming should not aim at merely emulating the abstraction from classical programming, but rather at extending the principles of high-level structured programming into the quantum domain~\cite{ying2024foundations,murillo2025qse}, making for a quantum-oriented paradigm~\cite{ali2023need}. 

In an effort towards increasing the abstraction level of quantum programming, this work introduces a conceptual framework that considers a quantum computation as a structure that evolves throughout the computation. To clarify, evolving means modifying the global quantum state by altering, individually or via composition, the distinct components that conform the structure. This structure is to be composed of one or several quantum registers, understood as ordered sets of qubits. Each register models what we denote as a computational construct, that is, an entity modeled (e.g., energy levels of a molecule or air flows in a weather model) within available computation resources. Then, operations are no longer thought at the level of quantum gates acting on individual qubits, but at that of quantum register as undivided entities. To illustrate this approach, several high-level syntax are formalized as transformations for composing registers following schemes from previously well-known operations. This formalization is enabled through the definition of an algebraic formalism for quantum programming, proposed by Ying et al.~\cite{ying2026laws}. In particular, a quantum satisfiability modulo theories (SMT) solver is formulated relying solely on them: once the transformation are defined, a quantum program can be designed without referring to quantum gates. In essence, the proposed approach strives to bring to quantum programming some methodological principles of high-level structured decomposition, composition, and refinement.

The main contributions of this work are:
\begin{itemize}
    \item Introduction of a conceptual framework that serves as the basis for thinking about quantum computation processes at a higher-level than that of individual qubits evolution.
    \item Formalization of well-known operations for the conceptual framework using high-level syntax and an algebraic formalism that bridges them to their low-level semantics.
    \item The design of a quantum SMT solver program based on the transformations formalized, illustrating their potential capability of advancing quantum program design.
\end{itemize}

The rest of the work is organized as follows. In Section~\ref{quantum:programming} an overview for the current state of abstraction on quantum programming is presented. Before outlining the specifics of this proposal, Section~\ref{algebra} provides an explanation about the algebraic formalism used and its associated laws. Section~\ref{proposal} describes in detail the conceptual framework and formalizes the semantics of high-level syntax. Program design for a quantum SMT solver as an application using the previously defined syntax is shown in Section~\ref{SMT}. In Section~\ref{RW}, a brief discussion is offered. Lastly, Section~\ref{conclusions} recapitulates on the work and several conclusions are drawn.

\section{Abstraction on Quantum Programming}\label{quantum:programming}
In general terms, abstraction can be defined as a mechanism that allows the expression of relevant details, leaving irrelevant ones hidden~\cite{liskov1974programming}. Formally, it is a homomorphic mapping that allows the functioning of a structure to be preserved while ignoring its lower-level implementation. Fewer details mean that the barrier to understanding how the process works is lowered. For computational processes, this also may translate into better verifiability, modularity, and scalability~\cite{clarke1994model,baldwin2000design}.

Current quantum programming is located at an abstraction level similar to that of assembler languages~\cite{wilkes1958preparation} in classical programming. This foundation has been mainly procured by leveraging on binary terms\footnote{We refer to the computational basis $\{ \ket{0},\ket{1}\}$ when talking about binary terms. However, abstraction achieved through quantum circuits is valid for any orthonormal basis of the complex Hilbert space $\mathbb{C}^2$ physically realizable.} through the quantum circuit paradigm~\cite{deutsch1989quantum,chiyao1993} integrated in classical programming languages~\cite{chakraborty2011quect}. This advancement enables programmers to operate quantum computers using quantum gates, which resemble logical gates from classical computing, instead of descending to the level of physical implementation (e.g., pulse modulation~\cite{krantz2019quantum}). This is the paradigm most current quantum programming languages work at~\cite{heim2020quantum}, such as \textit{Scaffold}~\cite{abhari2012scaffold}, \textit{Quipper}~\cite{green2013quipper}, \textit{Q\#}~\cite{svore2018q}, \textit{Pennylane}~\cite{bergholm2022pennyLane}, \textit{Qiskit}~\cite{qiskit2024},  or \textit{Cirq}~\cite{cirq2025} among others. Notwithstanding, it is worth mentioning high-level quantum programming languages, namely \textit{Silq}~\cite{bichsel2020silq}, \textit{Qrisp}~\cite{seidel2024qrisp,bock2025designing} or Qmod~\cite{vax2025qmod}. These languages are capable of declaring quantum data on a higher level of abstraction, however, there is not a suitable framework that properly uses these to develop a structured programming paradigm taking advantage of the quantum nature of operations.

As such, quantum programs are mainly designed at the level of quantum gates sequences to adapt to these languages. Whereas many quantum algorithms are based on evolutions that can be, to some extent, directly mapped into quantum gates sequences~\cite{cleve1998quantum, dalzell2023quantum}, this translation is not always straightforward, especially when the scheme is out of the scope of a complete algorithm (e.g., when the goal is to use algorithm parts rather than an entire routine). Even though it is possible to assemble operations into custom (higher-level) gates acting on sets of qubits following the intended scheme, they need to be manually specified gate by gate. This process requires programmers to understand the operations behavior and design the program as a sequence of quantum gates acting on individual qubits. Given this, specification becomes nontrivial and error-prone when problems turn increasingly more complex and grow in scale. At the same time, given such limitation, expressive capacity and creation of novel algorithms is hindered. 

Under the premise that classical programming historical development achieved success through abstraction increase~\cite{shaw2025revisiting}, seems reasonable to expect that quantum programming can benefit from it as well~\cite{di2024abstraction}. Particularly, given the complexity and counter-intuitive nature of quantum mechanics, abstraction promises to hasten the adoption process of this technology~\cite{shih2021increasing}. Classical programming ceased to be implemented in binary terms and evolved towards high-level languages based on abstract data types~\cite{liskov1974programming}. Currently, classical programmers can model real elements into abstractions inside a program and specify their interactions (e.g., object-oriented programming~\cite{black2013object}), completely disengaging algorithmic logic from physical implementation. Meanwhile, quantum programming lacks a  a proper model where quantum information can be used for designing programs in a similar way. Although some higher abstractions have been formalized in a quantum setting, such as quantum modular arithmetic~\cite{vedral1996quantum,draper2000addition} and circuit synthesis~\cite{yan2025quantum,goldfriend2025design}, a more versatile framework that treats quantum information as constructs that establish relations is yet to be introduced. Additionally, such abstraction must be aligned with the internal workings of a quantum computer as to not unintentionally violate its basic principles, leading to inappropriate or not feasible computations~\cite{zhao2025abstraction}.

Hence, this work aims towards establishing a quantum-oriented structured programming paradigm. This is achieved by regarding quantum computations as structures that can be modified via composition and refinement of quantum registers. However, it is important to note that the underlying implementation of such quantum programming paradigm may result in a more complex arrangement than what actual quantum computing hardware is able to handle. (That is one of the main reasons behind the disparity between research progress in both areas.) Yet deferring quantum programming innovation until quantum hardware architectures reach full operational maturity is a methodologically flawed proceeding.

\section{Algebraic Formalism}\label{algebra}
To properly establish the foundations of this conceptual framework and the semantics of the proposed transformations, it is appropriate to use an adequate formalism capable of exhibiting the quantum nature of the operations. The work by Ying et al.~\cite{ying2026laws} provides a formalism for treating transformations as structured constructs, extending Hoare's basic laws of classical programming~\cite{hoare1987laws} to the quantum domain, allowing for expressing low-level semantics as algebra. With this formalism, operations that influence entire registers can be modeled using high-level statements such as sequential compositions, quantum conditionals, or loops. It is particularly interesting due to its capability of verifying recursively defined quantum circuits and representing lower-level abstractions when operating on individual qubits~\cite{ying2024verification}. Therefore, it seems reasonable to argue that it provides appropriate tools for establishing verifiable programs at a higher level.

Although the formalism covers both quantum circuits and programs (i.e., provided with mid-circuit measurements), in this work we will only introduce the former. The reason being that the transformations defined below are all formalized as quantum circuits, as none of them present mid-circuit measurements. However, not all laws regarding quantum circuits are introduced, only those relevant for this work. (Refer to Section 3 and 4 of~\cite{ying2026laws} for a complete overview.)

\subsection{Quantum circuits}
We consider a set $\mathcal{QV}$ of quantum variables. Each quantum variable $q \in \mathcal{QV}$ represents a quantum system with an associated Hilbert space denoted by $\mathcal{H}_q$. A sequence $\overline{q}=q_1,\dots,q_n$ of distinct quantum variables is called a quantum register. That is, a quantum register consists of a quantum system composed of subsystems $q_1,\dots,q_n$ (typically, these subsystems can be thought as individual qubits), where its corresponding Hilbert space is given by $\bigotimes^n_{i=1}\mathcal{H}_{q_i}$. Moreover, sequence $\overline{q}_1, \dots, \overline{q}_N$ denotes the concatenation of $N$ registers. Additionally, we have a set $\mathcal{U}$ of unitary matrix constants. Each unitary matrix constant $U[\{q_i\}_{i=1}^m] \in \mathcal{U}$ denotes a unitary operation acting on the subspace spanned by a set of $m$ subsystems $\bigotimes_m \mathcal{H}_{q_m}$. Notice that for a given register $\overline{q}$ composed of $n$ subsystems, if $m=n$, then $U[\overline{q}]$ denotes a quantum operation over the whole register.

In accordance with this basic notation, quantum circuits $C \in \mathcal{QC}$ are defined such that:
\begin{equation}
    C ::= \mathbf{skip} \mid U[\overline{q}] \mid C_1;\dots;C_j \mid \mathbf{qif}[\overline{q}](\square_{i=1}^{d}\ket{\psi_i} \rightarrow C_i)\mathbf{fiq}.
\end{equation}

\noindent Here, $\mathbf{skip}$ denotes an identity operator. $U[\overline{q}]$ represents an operation $U$ acting on the register $\overline{q}$. Meanwhile, $C_1;\dots;C_j$ is a sequence of circuits (also presented in this work as $\mathbf{seq}_{l=1}^j C_l$ in order to improve readability). Moreover, $\mathbf{qif}$ is a quantum if-statement where $\{\ket{\psi_i}\}^d_{i=1}$ is an orthonormal basis of the Hilbert space associated to control register $\overline{q}$, and $\square_{i=1}^{d}$ denotes the branching over all possible control states with $C_i\ (i=1,\dots,d)$ illustrating different circuits acting on quantum variables outside register $\overline{q}$. 

\subsection{Algebraic laws for quantum circuits}
Once quantum circuits are defined, we can continue introducing several laws regarding them. These laws are mainly divided into two categories: quantum if-statement and sequential composition. To better describe these laws, we first introduce a different notation for quantum if-statements with two possible outcomes:
\begin{equation}\label{conditional:def}
\resizebox{\linewidth}{!}{$
    C_0 \xleftarrow{\ket{\psi_0}} q \xrightarrow{\ket{\psi_1}} C_1 \triangleq \mathbf{qif} \ [q] \ (\ket{\psi_0} \rightarrow C_0 \ \square \ \ket{\psi_1} \rightarrow C_1) \ \mathbf{fiq},
$}
\end{equation}

\noindent with $q$ denoting a single qubit with basis $\{\ket{\psi_0},\ket{\psi_1}\}$. It is worth noting that this also applies to a quantum register $\overline{q}$ acting as control whenever only two states from its corresponding basis are taken into account. If the quantum if-statement relies on more than two branches, this notation is not suitable.

\subsubsection{Laws of quantum if-statement}
Rules regarding a quantum conditional operator can be formalized as follows. When an unitary $U$ changes a basis $\ket{\psi_0}, \ket{\psi_1}$ to $\ket{\varphi_0}, \ket{\varphi_1}$:
\begin{equation}\label{math:basischange}
    C_0 \xleftarrow{\ket{\psi_0}} q \xrightarrow{\ket{\psi_1}} C_1 \equiv U^\dagger[q]; C_0 \xleftarrow{\ket{\varphi_0}} q \xrightarrow{\ket{\varphi_1}} C_1; U[q].
\end{equation}

\noindent Quantum conditionals satisfy the idempotence property:
\begin{equation}\label{idempotence}
    C \xleftarrow{\ket{\psi_0}} q \xrightarrow{\ket{\psi_1}} C \equiv C.
\end{equation}

\subsubsection{Laws of sequential composition}
Given a sequence of circuits, its composition can be rewritten as:
\begin{equation}\label{compose}
    U_1[\overline{q}]; U_2[\overline{q}] \equiv (U_2U_1)[\overline{q}],\ \text{if}\ U_2U_1\in\mathcal{U}.
\end{equation}

\noindent Regarding that quantum variables used throughout circuit $C$ as $qv(C)$, circuit sequences that satisfy $qv(C_1) \cap qv(C_2) = \emptyset$ obey the commutativity rule:
\begin{equation}\label{commutativity}
    C_1;C_2 \equiv C_2;C_1
\end{equation}

\noindent As such, if the composition is established in parallel:
\begin{equation}\label{parallel:sequence}
\begin{aligned}
    U_1[\overline{q}]; U_2[\overline{r}] \equiv U_2[\overline{r}]; U_1[\overline{q}] \equiv (U_1 \otimes U_2)[\overline{q}, \overline{r}],\\ \text{if}\ \overline{q}\cap\overline{r}=\emptyset\ \text{and}\ (U_1\otimes U_2)\in\mathcal{U}.
\end{aligned}
\end{equation}

\noindent By making use of Eq.~\eqref{idempotence}, the following equivalences for a sequence involving a quantum conditional can be obtained: 
\begin{align}\label{conditional1}
    C; (C_0 \leftarrow q \rightarrow C_1) &= (C; C_0) \leftarrow q \rightarrow (C; C_1), \\
    \label{conditional2}
    (C_0 \leftarrow q \rightarrow C_1); C &= (C_0; C) \leftarrow q \rightarrow (C_1; C).
\end{align}

\noindent Likewise, for a sequence comprising quantum conditionals with the same control conditions, its composition follows:
\begin{equation}\label{dist:cond}
\resizebox{\linewidth}{!}{$
    (C_0 \xleftarrow{\ket{\psi_0}} q \xrightarrow{\ket{\psi_1}} C_1); (D_0 \xleftarrow{\ket{\psi_0}} q \xrightarrow{\ket{\psi_1}} D_1) \equiv (C_0; D_0) \xleftarrow{\ket{\psi_0}} q \xrightarrow{\ket{\psi_1}} (C_1; D_1).
$}
\end{equation}

\section{Conceptual Framework}\label{proposal}
In this work, a conceptual framework for enabling a higher level of abstraction for designing quantum programs is proposed. The aim is centered around extending principles from structured programming to the quantum field. 

First, we start from considering quantum registers as operands that model computational constructs. In essence, a quantum program can be devised as a computable quantum state that evolves according to the transformations of quantum registers, given that these model computational constructs that present quantum traits. These transformations are operations that act on single quantum registers or that entangle two or more of them through composition operations. By performing a series (sequences) of semantically meaningful transformations following a (programming) strategy towards a solution, we are able to modify the global state to obtain the desired result. This directly contributes to the abstraction of the program: by realizing operations on sets of quantum registers as undivided entities instead of individual qubits, less components are instantiated and less details need to be specified.

The proposal continues by denoting how a quantum register for realizing logical evaluations is addressed. With this, it is possible to formalize a series of transformations that are used to generalize previously well-know operations among registers. These operations are based on quantum phase shifting due to logical constraint satisfaction~\cite{grover1996search,brassard2000quantum,szegedy2004quantum} and parallel evaluation~\cite{greenberger1989going,moore2001parallel}. 

High-level syntax based on these concepts is presented for increasing the abstraction, however, implementation of these transformations continues to take place at the level of quantum gates on individual qubits. Then, the algebraic formalism introduced before allows for demarcating how quantum information is handled and their associated low-level semantics. Nevertheless, programmers should continue to work with high-level syntax. As such, a suitable quantum programming language ought to be able to translate these transformations into operations on individual qubits for their physical implementation. For this reason, we specify both the functionality and implementation of the formalized abstractions. 

\subsection{Quantum registers as operands}\label{qregs}
A quantum register can be described as follows. Let $\mathcal{H}_{\overline{q}}$ be the complex Hilbert space associated with the quantum register $\overline{q}$, with dimension $D = \dim(\mathcal{H}_{\overline{q}})$. We define a complete orthonormal set $\{\ket{\psi_k}\}_{k=1}^{D}$ that constitutes an arbitrary basis for $\mathcal{H}_{\overline{q}}$. The register $\overline{q}$ is described as the superposition of basis elements as:

\begin{equation}
    \overline{q} := \sum_{k=1}^{D} c_k \ket{\psi_k},
\end{equation}

\noindent where complex coefficients $c_k\in \mathbb{C}$ encode the information embedded into each register. These coefficients can be decomposed into their probability amplitude and phase components according to $c_k = r_k e^{i\varphi_k}$. Term $r_k = |c_k| \in [0, 1]$ represents the probability amplitude, satisfying the global normalization condition $\sum_{k=1}^{D} |c_k|^2 = 1$; whereas $\varphi_k \in [0, 2\pi)$ denotes the relative phase.

Towards realizing a quantum-oriented structured programming, quantum registers are considers as operands of transformations realized through quantum computations (instead of individual qubits). These are able to represent abstract data on a higher level than binary terms following a type-specific encoding. In a sense, this can be extended so that quantum registers can model computational constructs. That is, a quantum register is capable of computationally model a construct by representing with its basis elements different states regarding that construct. For instance, basis elements of a quantum register with $D=4$ could each represent one of four different water levels in a bottle.

In consequence, it is possible operate with a quantum register that acts as a computational construct while hiding details about lower-level operations acting on their constituent qubits. That is, the system dynamics are describe in the  complex Hilbert space $\mathcal{H} = \bigotimes_{j} \mathcal{H}_{\overline{q}_j}$ with unitary operations defined as $\mathcal{U} : \mathcal{H} \rightarrow \mathcal{H}$, taking registers as single, undivided entities. Albeit higher in abstraction, this conceptual framework is to be engaged from two different perspectives: establishing single register transformations dependent on data encoding (i.e., altering a single quantum register based on the entity it models) or ascertaining operations that compose several registers. This work is mainly focused on formalizing composition operations that are not dependent on which computational construct is being encoded. Nonetheless, the internal structure of these transformations would vary depending on the abstract data, prompting distinct single register transformations (e.g., treating a quantum register as representing an integer or a string), even if the overall transformation scheme formalized is left invariant.

\subsection{Quantum registers for logical constraint transformations}
In order to evaluate complex logical relations in quantum registers, it is convenient to structure the Hilbert space associated with each register introducing a subspace dedicated to condition evaluation. Each operand register must contain ancilla\footnote{Henceforth, we will use the term \textit{ancillas} to refer to additional qubits involved throughout computation, and the term \textit{auxiliary} for those not needed but practical for optimization.} qubits that encode information about logical conditions placed upon the register. It is important to note that the register and ancillas involved in this operation will be entangled at the end of the operation, thus uncomputation is relevant. 

Considering an arbitrary register $\overline{q}$ composed of $m$ qubits whose Hilbert space can be factorized into $\mathcal{H}_{\overline{q}} = \mathcal{H}_{\text{val}} \otimes \mathcal{H}_{\text{anc}}$. Here, $\mathcal{H}_{\text{val}} \cong (\mathbb{C}^2)^{\otimes m-n}$ with $n$ the number of logical constrains enforced upon the register, contains the superposition of states which yield all register possible values. Meanwhile, $\mathcal{H}_{\text{anc}} \cong (\mathbb{C}^2)^{\otimes n}$ represents ancilla qubits that store information about the evaluation of certain logical propositions or conditions for each of the superposition states. Then, a valid register $\overline{q}$ for evaluating $n$ logical relations follows:

\begin{equation}\label{codification}
\overline{q} := \sum_{x=0}^{2^{m-1}-1} c_x \ket{\psi_x} \otimes \ket{f_1(\psi_x)}\otimes \dots \otimes  \ket{f_n(\psi_x)},
\end{equation}

\noindent where the set $\{\ket{\psi_x}\}$ constitutes an orthonormal basis that encodes data in the subspace $\mathcal{H}_{\text{val}}$, and $\ket{f_i(\psi_x)} \in \{\ket{0}, \ket{1}\}$ denotes an ancilla qubit that evaluates the boolean proposition $f_i$ over the set of values.

Note that this quantum register structure is thought for logical evaluation, and it generates entanglement between ancillas and the register. As such, ancillas are temporal (to be uncomputed) resources within this context, and transformations of other nature do not explicitly require appended ancillas.

\subsection{Abstraction: PhaseAND\texorpdfstring{$(\theta)$}{(theta)}}\label{abs:PhaseAND}
\begin{figure}[b]
\centerline{\includegraphics[width=0.55\textwidth]{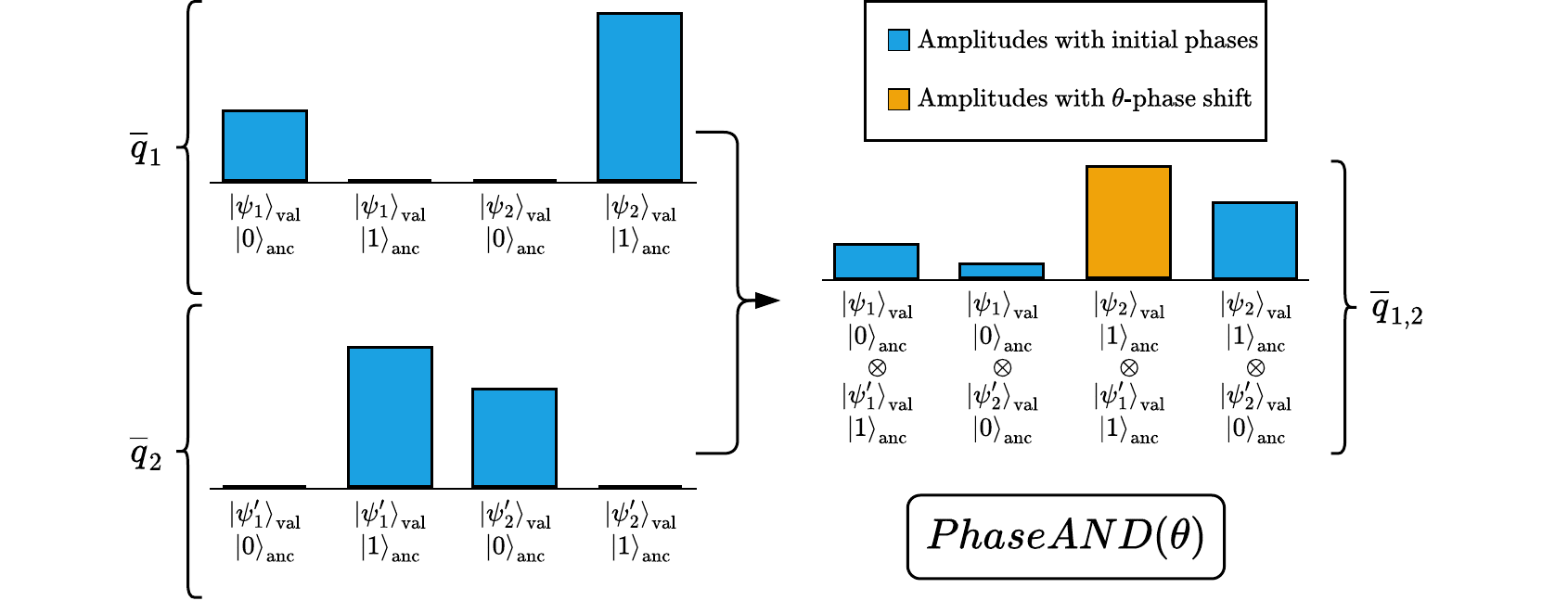}}
\caption{Representation of the $PhaseAND(\theta)$ operation functionality on two register with one constraint each. On the right, those states without amplitude contribution are suppressed for simplification. Bars height is an arbitrary depiction of amplitude absolute values while color is related to relative phase.}\label{fig:phaseAND}
\end{figure}

Within the given structure, the logical conjunction of $N$ conditions is the first operation formalized, namely $PhaseAND(\theta)$ (or $Ph_{\wedge,\theta}$ for simplicity). As dictated previously, each involved register must encode, at least, one constraint; thus, bi-partitioning every state of the superposition following ancilla evaluation. This transformation objective is to apply a $\theta$-phase shift, to the global system if and only if all conditions encoded on the involved registers are satisfied simultaneously (see Fig.~\ref{fig:phaseAND}). This is a composition operation where several computational constructs, each subjected to their own constraints, are arranged into a combined global state that applies phase shift if the logical conjunction of all constraints---their own and those of others---is satisfied. 

That being said, its high-level syntax follows a quantum if-statement as:
\begin{equation}\label{PhaseAND:eq}
\resizebox{\linewidth}{!}{$
    Ph_{\wedge,\theta}[\{\overline{q}_i\}^{N}_{i=1}] \equiv \mathbf{qif}\ \bigwedge_{i=1}^N \bigwedge_{j=1}^{n_i} f_{j,\overline{q}_i}\ \mathbf{then}\ Ph_\theta[\{\overline{q}_i\}^{N}_{i=1}]\ \mathbf{else}\ \mathbf{skip}\ \mathbf{fiq},
$}
\end{equation}

\noindent with $n_i$ the number of logical constraint the $i$-th register is subjected to, and $f_{j,\overline{q}_i}$ characterizes each constraint involved. Also, here $Ph_\theta$ is a circuit that applies a $\theta$-phase shift, and $\{\overline{q}_i\}^{N}_{i=1}$ denoting the set of involved registers. Involved register are entangled after applying the operation, whether or not they were entangled beforehand. Moreover, there is no restriction to the registers sizes and basis.

Eq.~\eqref{PhaseAND:eq} represents a high-level abstraction intended for a straightforward understanding, however, this representation needs to be translated into a quantum circuit for its physical implementation. Its corresponding realization can be implemented using a single auxiliary qubit. $Ph_{\wedge,\theta}$ creates an entangled system by performing a multi-controlled circuit on the auxiliary qubit where ancillas act as controls. Such circuit is $Ph_\theta$, and said circuit can be obtained through the gate sequence\footnote{Whereas $Ph_\theta$ represents a circuit that applies a $\theta$-phase shift to the whole state, $P_\theta$ denotes the quantum gate: $P_\theta = \left(\begin{smallmatrix} 1 & 0 \\ 0 & e^{i\theta} \end{smallmatrix}\right)$.} $X[q];P_\theta[q];X[q];P_\theta[q]$ acting on the auxiliary qubit $q$. In particular, circuit $Ph_\theta$ generates a global $\theta$-phase shift. Inherently, this phase shift is exclusively extended to the entangled system when controls are activated (i.e., in state $\ket{1}$).

However, in order to formalize operations without resorting to auxiliary qubits, we consider one of the involved ancillas as the target while the rest act as controls. Thus, ancillas associated to the set $\{{\overline{q}_i}\}^{N}_{i=1}$ perform a controlled operation acting on one of them, which is ruled out from the control. If all control ancillas are in state $\ket{1}$, a $P_\theta$ gate is applied to the target ancilla. That is, a multi-controlled $P_\theta$ gate acting on the ancilla qubits of involved registers. Then, high-level syntax from Eq.~\eqref{PhaseAND:eq}, selecting the last qubit appended as target, can be expressed with the algebraic formalism as:
\begin{equation}\label{qifPhaseAND}
\resizebox{\linewidth}{!}{$
    Ph_{\wedge,\theta}[\{\overline{q}_i\}^{N}_{i=1}] \equiv \mathbf{qif}\ [\{\overline{q}_i\}^{N}_{i=1}]\left( \bigotimes_{i=1}^{N} \square^{x_i} \ket{\psi_{x_i}}\ket{1}^{\otimes n_i} \rightarrow P_\theta[{\overline{q}_N}(t)] \right),
$}
\end{equation}

\noindent where $\square^{x_j}$ denotes the branching over all possible data states $\ket{\psi_{x_j}}$, ensuring the operation only depends on ancillas and $n_i$ (except the register related to the target ancilla, where $n_i=n_N-1$). Lastly, the circuit $P_\theta[{\overline{q}_N}(t)]$ indicates the application of a $P_\theta$ gate strictly on the target ancilla qubit $t$ of register $\overline{q}_N$. (The equivalence between applying the multi-controlled circuit $Ph_\theta$ on an auxiliary qubit and that of Eq.~\eqref{qifPhaseAND} is proven in Section~\ref{abs:compose}.) Of particular note is the possibility to specify this abstraction when only one constraint on one register is specified. This would imply placing a $P_\theta$ gate on the ancilla without controls.

Once the transformation $Ph_{\wedge,\theta}$ is formalized, we can obtain $PhaseNAND(\theta)$ (or $Ph_{\neg(\wedge),\theta}$). This can be achieved by adding an opposite global phase shift from that of $Ph_{\wedge,\theta}$, i.e, applying circuit $Ph_{-\theta}$ to any qubit of the entangled state:
\begin{equation}
\resizebox{\linewidth}{!}{$
\begin{aligned}
    Ph_{\neg(\wedge),\theta}[\{\overline{q}_i\}^{N}_{i=1}] & \equiv Ph_{\wedge,-\theta}[\{\overline{q}_i\}^{N}_{i=1}]; Ph_\theta[\{\overline{q}_i\}^{N}_{i=1}] \\
    & \equiv \mathbf{qif}\ \bigwedge_{i=1}^N \bigwedge_{j=1}^{n_i} f_{j,\overline{q}_i}\ \mathbf{then}\ \mathbf{skip}\ \mathbf{else}\ Ph_\theta[\{\overline{q}_i\}^{N}_{i=1}]\ \mathbf{fiq} \\
    & \equiv \mathbf{qif}\ \neg\left(\bigwedge_{i=1}^N \bigwedge_{j=1}^{n_i} f_{j,\overline{q}_i}\right)\ \mathbf{then}\ Ph_\theta[\{\overline{q}_i\}^{N}_{i=1}]\ \mathbf{else}\ \mathbf{skip}\ \mathbf{fiq}.
\end{aligned}
$}
\end{equation} 

\noindent Take into account that these abstractions are equivalent up to a global phase factor. Ergo, choosing one or another may yield the same results depending on the program design.

\subsection{Abstraction: PhaseOR\texorpdfstring{$(\theta)$}{(theta)}}\label{abs:PhaseOR}

\begin{figure}[t]
\centerline{\includegraphics[width=0.55\textwidth]{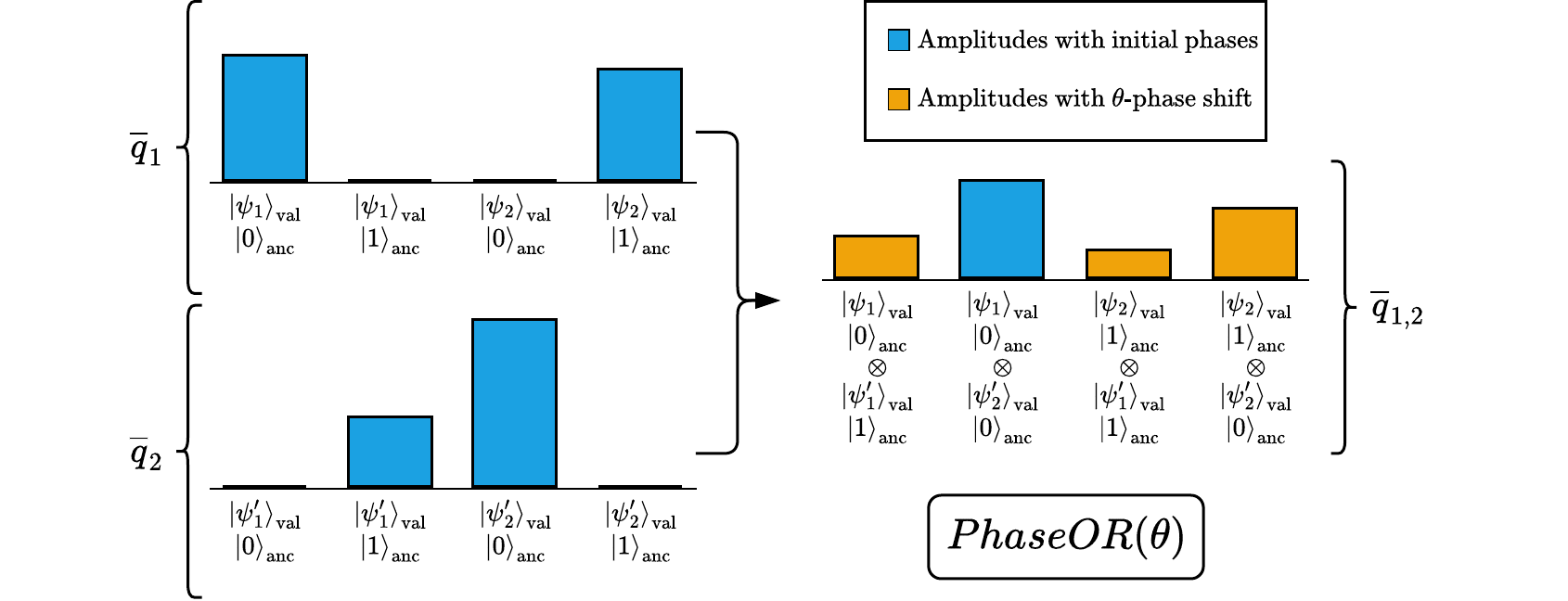}}
\caption{Representation of the $PhaseOR(\theta)$ operation functionality on two register with one constraint each. On the right, those states without amplitude contribution are suppressed for simplification. Bars height is an arbitrary depiction of amplitude absolute values while color is related to relative phase.}\label{fig:phaseOR}
\end{figure}

Following the same methodology, we can formalize the operation $PhaseOR(\theta)$ (or $Ph_{\vee,\theta}$). Analogously as with $Ph_{\wedge,\theta}$, the registers must follow Eq.~\eqref{codification} and after the operation an entangled system is obtained. This operation acts in the same manner as $Ph_{\wedge,\theta}$ but evaluates a disjunction instead (see Fig.~\ref{fig:phaseOR}). The associated quantum conditional is not as straightforward as for the conjunction, which only required control and target ancillas in state $\ket{1}$. Here, there are several possible combinations that satisfy the logical relation. For this reason, we infer its structure from that of $Ph_{\wedge,\theta}$. In particular, we exploit the relation between disjunction and conjunction such that $\neg\left(\bigvee f_j\right) \equiv \bigwedge \neg f_j$. That is, logical $NOR$ is equivalent to logical $AND$ on negated inputs.

Hence, the abstraction $PhaseNOR(\theta)$ (or $Ph_{\neg(\vee),\theta}$) can be constructed from $Ph_\wedge$ in a sequence along the input negations:
\begin{equation}
\resizebox{\linewidth}{!}{$
\begin{aligned}\label{qif:PhaseOR}
    Ph_{\neg(\vee),\theta}[\{\overline{q}_i\}^{N}_{i=1}] & \equiv \mathbf{seq}_{i=1}^N\mathcal{X}_{\overline{q}_i}; Ph_{\wedge,\theta}[\{\overline{q}_i\}^{N}_{i=1}];\mathbf{seq}_{i=1}^N\mathcal{X}_{\overline{q}_i},
    \\
    & \equiv\mathbf{qif}\ \neg\left(\bigvee_{i=1}^N \bigvee_{j=1}^{n_i} f_{j,\overline{q}_i}\right)\ \mathbf{then}\ Ph_\theta[\{\overline{q}_i\}^{N}_{i=1}]\ \mathbf{else}\ \mathbf{skip}\ \mathbf{fiq},
\end{aligned}
$}
\end{equation}

\noindent with $\mathbf{seq}_{i=1}^N\mathcal{X}_i$ being equal to a parallel sequence, as expressed by Eq.~\eqref{parallel:sequence}, of $N$ distinct $\mathcal{X}$ circuits. Each $\mathcal{X}_{\overline{q}_i}$ circuit is equivalent to $(\mathbbm{1}^{m_i} \otimes X^{n_i})[\overline{q}_i]$, where Pauli-$X$ gates are solely applied on ancillas and $m_i$ indicates the number of qubits register $\overline{q}_i$ is composed of. By applying $\mathbf{seq}_{i=1}^N\mathcal{X}_{\overline{q}_i}$ again ($\mathcal{X}$ is hermitian) after $Ph_{\wedge,\theta}$, the original codification is recovered, following Bennett's principle~\cite{bennett1973reversibility}. Then, $Ph_{\vee,\theta}$ can be deduced as a sequence composed of $Ph_{\neg(\vee),\theta}$ and $Ph_{-\theta}$:
\begin{equation}\label{PhaseOR:eq}
\resizebox{\linewidth}{!}{$
\begin{aligned}
    Ph_{\vee,\theta}[\{\overline{q}_i\}^{N}_{i=1}] & \equiv Ph_{\neg(\vee),-\theta}[\{\overline{q}_i\}^{N}_{i=1}]; Ph_\theta[\{\overline{q}_i\}^{N}_{i=1}] \\
    & \equiv \mathbf{qif}\ \bigvee_{i=1}^N \bigvee_{j=1}^{n_i} f_{j,\overline{q}_i}\ \mathbf{then}\ Ph_\theta[\overline{q}_1, \dots, \overline{q}_N]\ \mathbf{else}\ \mathbf{skip}\ \mathbf{fiq}.
\end{aligned}
$}
\end{equation}

\subsection{Abstraction: Reflection}\label{abs:REFLECTION}
Stemming from these logical transformations, it is possible to formalize a more complex one. This operation represents a reflection across a certain subspace $\ket{\psi_k}$. Mathematically, it can be expressed as $2\Pi_k-\mathbbm{1}$, with $\Pi_k = \ket{\psi_k}\bra{\psi_k}$ the projector over subspace spanned by $\ket{\psi_k}$. From considering a basis change such as $U_k\Pi_0U_k^\dagger=\Pi_k$, where $\Pi_0=\ket{0\dots0}\bra{0\dots0}$, the operation can be rewritten as $U_k(2\Pi_0-\mathbbm{1})U_k^\dagger$. Operator $U_k$ represents an arbitrary state initialization following $U_k\ket{0...0} = \ket{\psi_k}$. The corresponding circuit will vary depending on the selected subspace. Conversely, $2\Pi_0-\mathbbm{1}$ denotes a well-defined operation that can be formalized. It represents a $\pi$-phase shift acting on all states except $\ket{0\dots0}$. 

Notice that this last operation can be obtained by performing $Ph_{\neg(\wedge),\theta}$, taking $\theta=\pi$ and constraints that activate one ancilla exclusively for initial state $\ket{0\dots0}$ on each involved register. This initial state is independent of initialization, which is determined by the construction of operator $U$. A single register reflection is expressed as:
\begin{equation}\label{math:reflec}
\begin{aligned}
    REFL_{\Pi_0}[\overline{q}] & \equiv C_{\Pi_0}[\overline{q}];Ph_{\neg(\wedge),\pi}[\overline{q}];C_{\Pi_0}^\dagger[\overline{q}] \\
    & \equiv \mathbf{qif}\ \Pi_0\ \mathbf{then}\ \mathbf{skip}\ \mathbf{else}\ Ph_\pi[\overline{q}]\ \mathbf{fiq},
\end{aligned}
\end{equation}

\noindent where $C_{\Pi_0}$ denotes a circuit that acts on the ancilla conditioned by projector $\Pi_0$. Circuit $C_{\Pi_0}$ can be achieved by placing a negated multi-controlled operation acting on the ancilla with all value qubits from the register as controls. Correspondingly, a generalization acting on $N$ registers follows:
\begin{equation}\label{math:refl:0}
\resizebox{\linewidth}{!}{$
\begin{aligned}
    REFL_{\Pi_0}[\{\overline{q}\}_{i=1}^N] & \equiv \mathbf{seq}_{i=1}^N C_{\Pi_{0,\overline{q}_i}}^\dagger;Ph_{\neg(\wedge),\pi}[\{\overline{q}\}_{i=1}^N];\mathbf{seq}_{i=1}^N C_{\Pi_{0,\overline{q}_i}} \\
    & \equiv \mathbf{qif}\ \bigwedge_{i=1}^N \Pi_{0,\overline{q}_i}\ \mathbf{then}\ \mathbf{skip}\ \mathbf{else}\ Ph_\pi[\{\overline{q}\}_{i=1}^N]\ \mathbf{fiq},
\end{aligned}
$}
\end{equation}

\noindent taking $\mathbf{seq}_{i=1}^N C_{\Pi_{0,\overline{q}_i}}$ as a parallel sequence of $N$ circuits $C_{\Pi_{0,\overline{q}_i}}$ acting each on its associated register $\overline{q}_i$, where $\Pi_{0,\overline{q}_i}$ denotes the corresponding projector for state $\ket{0\dots 0}$.

In consequence, regarding $C_{k}$ as the circuit that performs state initialization given by operator $U_k$, the abstraction for reflection across subspace $\ket{\psi_k}$ acting on one register follows:
\begin{equation}
\begin{aligned}
    REFL_{\Pi_k}[\overline{q}] & \equiv C_k^\dagger[\overline{q}];REFL_{\Pi_0}[\overline{q}];C_k[\overline{q}]\\
    & \equiv \mathbf{qif}\ \Pi_k\ \mathbf{then}\ \mathbf{skip}\ \mathbf{else}\ Ph_\pi[\overline{q}]\ \mathbf{fiq}.
\end{aligned}
\end{equation}

\noindent And its generalization acting on $N$ distinct register can be derived as:
\begin{equation}\label{math:refl:k}
\resizebox{\linewidth}{!}{$
\begin{aligned}
    REFL_{\Pi_{k_i}}[\{\overline{q}_i\}^{N}_{i=1}] & \equiv \mathbf{seq}_{i=1}^NC_{k_i}^\dagger;REFL_{\Pi_0}[\{\overline{q}_i\}^{N}_{i=1}];\mathbf{seq}_{i=1}^N C_{k_i} \\
    & \equiv \mathbf{qif}\ \bigwedge_{i=1}^N \Pi_{k,\overline{q}_i}\ \mathbf{then}\ \mathbf{skip}\ \mathbf{else}\ Ph_\pi[\{\overline{q}\}_{i=1}^N]\ \mathbf{fiq},
\end{aligned}
$}
\end{equation}

\noindent where each circuit $C_{k_i}$ realizes a distinct operator $U_{k_i}$ associated to each register $\overline{q}_i$. Each circuit $C_{k_i}$ will have a different structure depending on how the target state is modeled for distinct computational constructs.

This operation can be extended further. Take as an example a case where reflection needs to be performed over the orthogonal subspace $\ket{\psi_k}^\perp$. Given the structure of Eq.~\eqref{math:refl:k}, this can be achieved by modifying the quantum if-statement control with its negation. That is, the middle-sequence component of Eq.~\eqref{math:refl:0} applies $Ph_{\wedge,\pi}$ instead. Furthermore, consider another instance where strict inversion is not suitable but an arbitrary phase $\theta$ is (e.g., fixed-point quantum search~\cite{grover2005fixed,yoder2014fixed}). By adjusting Eq.~\eqref{math:refl:k} with a $Ph_\theta$ circuit and Eq.~\eqref{math:refl:0} with $Ph_{\wedge,\theta}$, this operation can be attained.

Reflection is a fundamental part of many algorithms such as quantum amplitude amplification~\cite{brassard2000quantum} or Szegedy quantum walks~\cite{szegedy2004quantum}, both relying on their iterative application. In fact, it  is a concrete case of a phase-shift operator over a certain subspace. If extended, following quantum singular value transformation~\cite{gilyen2019quantum,martyn2021grand}, schemes that apply phase-shift operators could be generalized as polynomial transformations (also used in Hamiltonian simulation~\cite{low2019hamiltonian}). Albeit in its early stages, this work addresses transformations that may prove useful across diverse areas and could benefit from a quantum structured programming (e.g., quantum machine learning~\cite{rebentrost2014quantum, biamonte2017quantum}, solving quantum linear systems problems~\cite{dervovic2018quantum}). 

\subsection{Abstraction: Synchronization}\label{abs:SYNC}
In parallel to the previously defined transformations, we address a state initialization transformation among registers that model the same computational construct. This operation, called Synchronization (or $SYNC$), stems from Greenberger-Horne-Zeilinger (GHZ) states~\cite{greenberger1989going}. In them, more than two entangled particles exhibit the same behavior. Mathematically, for a system of $n$ qubits: $(\ket{0}^{\otimes n}+\ket{1}^{\otimes n})/\sqrt{2}$. It is straightforward to acknowledge the reason behind the proposed name. If we establish a GHZ-like state among qubits in the same position over distinct registers, we obtain registers that effectively map the same values in parallel. The register states are not copied, as it is prohibited by the \textit{no-cloning theorem}~\cite{wootters1982single}, but their qubit values are mapped onto other registers:
\begin{equation}
\resizebox{\linewidth}{!}{$
    \sum_k^{D} \left(c_k\ket{\psi_k}\right)\otimes\ket{0\dots0}\otimes\cdots\otimes\ket{0\dots0} \xrightarrow{SYNC} \sum_k^D \left(c_k\ket{\psi_k}\otimes\ket{\psi_k}\otimes\cdots\otimes\ket{\psi_k}\right),
$}
\end{equation}

\noindent where $\{\ket{\psi_k}\}_k^D $ conform a basis for a complex Hilbert space of dimension $D$. Note that this holds if and only if all registers are the same in size, and exactly one of them encodes information in an arbitrary manner while the rest are all initialized in state $\ket{0\dots0}$. Based on how GHZ-like states are obtained, we can formalize this initialization abstraction as series of  controlled operations with a single control qubit each.

Consider that $N$ registers are involved and they are each composed of $m$ qubits. (When coupled with logical abstractions, $SYNC$ subject registers are not intended to include ancillas.) Each controlled operation is executed bit-wise: the $i$-th qubit of the control register directly targets the $j$-th qubit of a different register. This operation is iterated over all registers and for all qubits within the control register. Its high-level syntax is:
\begin{equation}\label{Sync}
SYNC[\{\overline{q}_i\}^{N}_{i=1}] \equiv \overline{q}_1 \mapsto \overline{q}_i\ \mathbf{for}\ i \in [2,N],
\end{equation}

\noindent with $i\in \mathbbm{Z}$, register $\overline{q}_1$ initialized in an arbitrary manner prior to the operation, and $\overline{q}_{i,j}$ denoting the $j$-th qubit on register $\overline{q}_i$. The corresponding algebraic formalism for low-level is:
\begin{equation}
\mathbf{seq}_{i=2}^N\left(\mathbf{seq}_{j=1}^m \mathbf{qif}\ \overline{q}_{1,j}\ \mathbf{then}\ X[\overline{q}_{i,j}]\ \mathbf{else}\ \mathbf{skip}\ \mathbf{fiq}\right).
\end{equation}

As a matter of fact, it is trivial to extend Eq.~\eqref{Sync} for the case of uniform superposition initialization of register $\overline{q}_1$. Encoding uniform superposition can be achieved by applying Hadamard gates in every qubit of a register initialized in state $\ket{0\dots0}$. As such, it can be obtained as a sequence composed of Hadamard gates on all qubits of register $\overline{q}_1$ and $SYNC$:
\begin{equation}
    SYNC_{\text{unif}}[\{\overline{q}_i\}^{N}_{i=1}]\equiv\bigotimes^{m}H[\overline{q}_1];SYNC[\{\overline{q}_i\}^{N}_{i=1}],
\end{equation}

\noindent with $m$ the number qubits within register $\overline{q}_1$. 

\subsection{Sequence optimization and verification}\label{abs:compose}
Regarding logical transformations, it has been demonstrated that new operations can be formalized based on previously defined ones as reflection has been constructed upon $PhaseNAND(\pi)$. In a sense, this is possible due to the composability of logical relations. Even so, it seems feasible to further construct more complex abstractions using those formalized beforehand. What is more, the algebraic formalism by Ying et al.~\cite{ying2026laws} not only has the capability of formalizing new transformations but, presumably, is also able to optimize certain sequences into simpler ones and verify equivalences. Next, we present an example on each of these capabilities.

\subsubsection{Optimization} 
Following the rules presented in Section~\ref{algebra}, sequences can be rewritten and a new sequence that may involve lower implementation costs can be obtained. For instance, a routine composed of sequence $(Ph_{\wedge,\theta}Ph_{\neg(\vee),\theta'}Ph_{\wedge,\theta''})[\{\overline{q}_i\}^{N}_{i=1}]$, can be optimized. Such optimization, owed to the algebraic characterization, is proven below. 

First, by illustrating the sequence via conditional notation as presented in Eq.~\eqref{conditional:def}:
\begin{equation}
    P_{\theta'}[\overline{q}_N(t)] \leftarrow [\{\overline{q}_i\}^{N}_{i=1}] \xrightarrow{\bigotimes_{i=1}^{N} \square^{x_i} \ket{\psi_{x_i}}\ket{1}^{\otimes n_i}} (P_{\theta}P_{\theta''})[\overline{q}_N(t)],
\end{equation}

\noindent with the conditional right side applied when all control ancillas are in state $\ket{1}$, and the left for all ancillas in state $\ket{0}$. Knowing that phase gates are additive ($P_\theta P_{\theta''}=P_{\theta+\theta''}$):
\begin{equation}\label{eq:equal}
\resizebox{\linewidth}{!}{$
    P_{\theta'}[\overline{q}_N] \leftarrow [\{\overline{q}_i\}^{N-1}_{i=1}] \rightarrow (P_{\theta+\theta''})[\overline{q}_N] \equiv (Ph_{\neg(\vee),\theta'}Ph_{\wedge,\theta+\theta''})[\{\overline{q}_i\}^{N}_{i=1}],
$}
\end{equation}

\noindent Assuming arbitrary finely-tuned phase modulation~\cite{mckay2017efficient}, this sequence has one less transformation and, in consequence, lesser implementation costs are achieved. 

\subsubsection{Verification}
Likewise, the algebraic framework serves as a tool for verifying equivalences between circuits, that is, their semantics are equivalent. In Section~\ref{abs:PhaseAND}, it was implied that a multi-controlled circuit $Ph_\theta$ targeting an auxiliary qubit was equivalent to a multi-controlled $P_\theta$ gate acting on the ancillas. To illustrate this equivalence we use conditionals definition.

From Eq.~\eqref{conditional:def} and for the computational basis, semantics regarding a single qubit conditional follow\footnote{Symbolically, $[\![\textit{syntax}]\!] \equiv \textit{semantics}$.}:
\begin{equation}\label{qubit:conditional}
    [\![U'\leftarrow q \rightarrow U ]\!] = \ket{0}\bra{0}_q\otimes U' + \ket{1}\bra{1}_q\otimes U.
\end{equation}
Assuming a setting with two ancilla qubits for simplicity, a multi-controlled circuit $Ph_\theta$ acting on an auxiliary qubit $q_3$:
\begin{equation}\label{verif1}
    q_1 \rightarrow \left(q_2 \rightarrow (X[q_3];P_\theta[q_3];X[q_3];P_\theta[q_3])\right),
\end{equation}

\noindent with the equivalent circuit for a controlled $P_\theta$ gate being:
\begin{equation}\label{verif2}
    q_1 \rightarrow P_\theta[q_2],
\end{equation}
with semantics:
\begin{equation}\label{verif3}
\resizebox{\linewidth}{!}{$
    [\![q_1 \rightarrow P_\theta[q_2]]\!] \equiv \ket{0}\bra{0}_{q_1}\otimes \mathbbm{1} + \ket{1}\bra{1}_{q_1}\otimes (\ket{0}\bra{0}_{q_2} + e^{i\theta}\ket{1}\bra{1}_{q_2}).
$}
\end{equation}

Arising from the fact that the controlled circuit in Eq.~\eqref{verif1} imposes a global $\theta$-phase shift, it is tantamount to operator $e^{i\theta}\mathbbm{1}$. Coupled with the conditional definition from Eq.~\eqref{qubit:conditional}, the semantics of Eq.~\eqref{verif1} are equivalent to:
\begin{equation}
\resizebox{\linewidth}{!}{$
\begin{aligned}
    & \equiv \ket{0}\bra{0}_{q_1}\otimes \mathbbm{1} + \ket{1}\bra{1}_{q_1}\otimes (\ket{0}\bra{0}_{q_2}\otimes \mathbbm{1} + \ket{1}\bra{1}_{q_2}\otimes e^{i\theta}\mathbbm{1}) \\
    & \equiv \ket{0}\bra{0}_{q_1}\otimes \mathbbm{1} + \ket{1}\bra{1}_{q_1}\otimes \left((\ket{0}\bra{0}_{q_2}+ e^{i\theta}\ket{1}\bra{1}_{q_2})\otimes \mathbbm{1}\right) \\
    & \equiv [\![q_1 \rightarrow P_\theta[q_2]]\!].
\end{aligned}
$}
\end{equation}

\subsection{Implementation costs}
Although this work is mainly focused on providing a high-level abstraction framework provided with several formalized operations, it is also important to account for their related implementation costs. For instance, $PhaseAND(\theta)$ (and the rest of logical transformations) imply a multi-controlled gate with $N-1$ controls. It has been proved that such construction can be simulated in terms of $\Theta(N^2)$ basic operations~\cite{barenco1995elementary}, and has since been improved to linear-depth~\cite{da2022linear}. However, this cost can be further reduced by leveraging on auxiliary qubits~\cite{balauca2022efficient,nie2024quantum,rosa2025optimizing}. Operations were previously defined so that they would be formally self-contained within the registers without involving external resources. Despite the selected approach for formalization, inclusion of auxiliary qubits may prove beneficial, although such inclusion ought to be program design-independent. In the case of reflection, implementation cost will be tied to the computational construct modeled and the selected subspace. As for synchronization, the sequence of controlled operations grows linearly with the number of registers (distinct control and target operations can be applied in parallel). In this case, using a linear amount of auxiliary qubits with the number of registers allows for reducing the complexity to logarithmic depth by performing fan-out computations~\cite{moore2001parallel}. Note that these complexities are presented assuming an \textit{all-to-all} connectivity, and they would need to be adapted for topologies that do not support it.

Regarding circuit width, a regime where the number of qubits used should not hinder the implementation as the proposal is intended for large-scale problems. Thereby, size do not pose a problem for as long as it is limited by available quantum resources.

\section{Application: Quantum SMT Solver}\label{SMT}
A suitable abstraction is one that, apart from hiding irrelevant details, serves as a useful piece that assists in solving a wide variety of problems. On account of this, the transformations formalized in Section~\ref{proposal} are chosen due to their applicability in several algorithms. 

With this in mind, an example of such applicability is shown; however, take into account that program design using this framework is limited to programs that rely solely on formalized operations. One such a case is Grover's search algorithm~\cite{grover1996search}. Following this algorithm's scheme, a quantum SMT solver can be addressed, based on previous work by Lin et al.~\cite{lin2024parallel}, using the operations from above. This solver accounts for solving logical formulae where each clause is encoded as a logical constraint restricted by a certain theory signature. In this case, the validation is conducted with clauses that follow relations from modular linear integer arithmetic; however, the selection of constructs is not restricted to modulo integers: an SMT solver can be understood as a combination of constructs that meet certain criteria. Given this, the algorithm design based on this conceptual framework serves as a structured scheme to any such problem, disregarding the internal data constructs may represent. 

\subsection{SMT formulation}
Let $x = (x_1, x_2, \dots, x_n)$ denote a set of variables. A formula $\mathcal{F}$, which is a boolean combination of $N$ clauses, in conjunctive normal form (CNF) is written as:
\begin{equation}
\mathcal{F}(x) = f_1(x) \land f_2(x) \land \dots \land f_N(x),
\end{equation}

\noindent if there exists an assignment $x^\ast$ such that $\mathcal{F}(x^\ast) = \textit{True}$, it is said the formula is satisfiable under $x^\ast$. In this notation, each clause $f_j$ is a disjunction of literals:
\begin{equation}
f_j(x) = \bigvee_{k} \ell_{j,k},
\ with \
\ell_{j,k} \in \{\varphi_{j,k}(x), \neg \varphi_{j,k}(x)\},
\end{equation}

\noindent with $k$ the index of each literal in the $f_j$ clause and $\varphi_{j,k}(x)$ a predicate (also called atom) of the form $\varphi_{j,k}(x) \;\equiv\; t_1(x) \;\bowtie\; t_2(x)$, where $t_1(x)$ and $t_2(x)$ are terms constructed from variables and function symbols and $\bowtie$ represents predicate symbols. These variables and symbols must be allowed by the respective theory signature $\Sigma$, which can be thought of as its \textit{alphabet}. For instance, in the interpretation for linear integer arithmetic ($\mathcal{LIA}$), predicate symbols satisfy $\bowtie \; \in \{=,\neq,>,<,\geq,\leq\}$; whereas function symbols are restricted to the set $\{+, -\}$\footnote{For a more profound understanding of SMT and its formalism, refer to~\cite{bradley2007calculus,de2011satisfiability,barrett2018satisfiability}.}.

Nonetheless, this is not the only manner in which SMT problems are addressed. A boolean combination may not be in a conjunctive normal form, but in a disjunctive normal form (DNF). Here, clauses are connected by conjunctions and each clause is a conjunction of literals. An SMT formula is then a boolean combination of atoms, forming a logical structure that abstracts theory-specific constraints into discrete boolean variables. Thereby, separating the logical structure from the underlying theory semantics.

\subsection{Oracles}
Encoding of information inside registers has been mentioned several times; despite this, no specification on how these encoding are achieved has been given thus far. Mainly, because encoding will depend on the data associated with a construct. Nevertheless, next is the manner in which encoding can be generally expressed for evaluating logical predicates. (This is the other perspective where this conceptual framework is engaged at: single-register transformations dependent on data encoding; still, it will not be further explained beyond oracle introduction in this work.) 

These encodings are intended for registers that follow the form of Eq.~\eqref{codification}. Where, as implied before, an ancilla evaluates a predicate for each value in superposition. Then, oracles are used as the components that enable these data encodings. In quantum computing, an oracle can be defined as a \textit{black box} that evaluates a certain function $f$ ensuing:
\begin{equation}
O_f \ket{\psi_x}\ket{\psi_y} = \ket{\psi_x}\ket{\psi_y\oplus f(\psi_x)},
\end{equation}

\noindent which modifies the state amplitudes following $f$ assessment. An oracle for function $f$ acting on register $\overline{q}_i$ can be expressed using the algebraic formalism as $O_{f}[\overline{q}_i]$.

\subsection{Algorithm}
A quantum approach for the SMT problem allows searching the assignments space in superposition, amplifying the amplitudes of those satisfying the constraints while suppressing the rest. In particular, the scheme designed by Lin et al.~\cite{lin2024parallel} relies on parallel evaluation of distinct clauses for solving the satisfiability (SAT) problem (where SMT is a generalization of SAT with variables that are not exclusively boolean). In alignment with its design, an extension for SMT that relies on the transformations proposed in this work is achieved. 

\begin{algorithm}[!b]
\caption{Quantum SMT solver program via Grover's search for multiple constructs synchronized as an algebraic sequence}
\label{alg:quantum_smt_ghz_partitioned}
\begin{algorithmic}[1]
\REQUIRE Formula $\mathcal{F}$ with $N$ clauses over $m$ constructs. Let $S_i \subseteq \{1, \dots, N\}$ be the subset of clause indices depending on construct $x_i$. Let $\overline{q}_{S_i}$ denote the corresponding subset of quantum registers.
\ENSURE Measurement yields a tensor product solution state $\ket{\psi_s} = \bigotimes_{i=1}^m \ket{\psi_s(x_i)}$ satisfying $\mathcal{F}$.

\STATE \textbf{Initialize:} 
\STATE $\overline{q}_{S_i} \leftarrow SYNC_{\text{unif}}[\overline{q}_{S_i}], \ \forall \ i \in \{1, \dots, m\}$

\STATE Calculate number of iterations $k$
\FOR{$l = 1$ \textbf{to} $k$}

    \STATE \COMMENT{\textit{Parallel clause evaluation}}
    \FOR{\textbf{each} register $j \in \{1, \dots, N\}$} 
        \STATE $\overline{q}_j \leftarrow O_{f_j}[\overline{q}_j]$ \COMMENT{\textit{Evaluate clause $f_j$ into $\overline{q}_j$ ancilla}}
    \ENDFOR

    \STATE \COMMENT{\textit{Global $\pi$-phase shift upon satisfaction of $\mathcal{F}$}}
    \STATE $\{\overline{q}_i\}_{i=1}^N \leftarrow Ph_{\wedge,\pi}[\{\overline{q}_i\}_{i=1}^N]$ (CNF)  
    \STATE \hspace{53pt} (or $Ph_{\vee,\pi}$ for DNF) 

    \STATE \COMMENT{\textit{Uncomputation}}
    \FOR{\textbf{each} register $j \in \{1, \dots, N\}$} 
        \STATE $\overline{q}_j \leftarrow O_{f_j}^\dagger [\overline{q}_j]$ \COMMENT{\textit{Clear ancillas}}
    \ENDFOR

    \STATE \COMMENT{\textit{Amplitude amplification via REFL}}
    \STATE $\{\overline{q}_i\}_{i=1}^N  \leftarrow  REFL_{\Pi_{k_i}}[\{\overline{q}_i\}_{i=1}^N ]$
    \STATE \COMMENT{\textit{$REFL_{\Pi_0}$ on construct representative registers}}
    \STATE ($\mathbf{seq}_{j \in S_i} C_{k_j} \equiv SYNC_{\text{unif}}[\overline{q}_{S_i}],\ \forall i \in \{1,\dots,m\}$)
\ENDFOR

\RETURN Measure one representative register $\overline{q}_{r}$ ($r\in S_i$) for each $i \in \{1, \dots, m\}$ to obtain $\ket{\psi_s}$
\end{algorithmic}
\end{algorithm}

By employing parallelization a computational construct exhibiting superposition is simultaneously evaluated for different clauses. This instance is especially relevant when trading-off increased width presents an advantage against concatenating evaluations of the same register. States satisfying all clauses are marked with a $\pi$-phase shift. Then, by performing iterations according to Grover's search~\cite{grover1996search}, it is possible to obtain solution states for the problem. Algorithm~\ref{alg:quantum_smt_ghz_partitioned} details the program design for a formula $\mathcal{F}$ that may depend on several construct (each variable denotes a distinct construct) where each construct is subjected to synchronization for evaluating a single clause per register involved. In case evaluation concatenation depth increase is preferred against width, registers should not be subjected to synchronization. Algorithm is conceived following Lin et al.~\cite{lin2024parallel}, thus synchronization is chosen; however, choosing whether synchronization is advantageous or not should be handled internally by a suitable programming language, not part of the high-level design. Notice that $REFL_{\Pi_0}$ can be applied to only one of the registers modeling the same construct, as it would be a redundant operation due to previous disentanglement brought about by $SYNC_{\text{unif}}^{\dagger}$. 

\subsection{Validation}
A validation of the designed quantum program is performed on Qiskit (v2.3.1) simulation environment~\cite{qiskit2024} by implementing the corresponding quantum program~\cite{chamizozenodo}. Transformations are generalized \textit{ad hoc} for the specific setting as this programming language resorts to individual gate specification. As such, a more adequate quantum programming language for future works should not rely on this low-level implementation.

Each register encodes a distinct clause using oracles based on modular arithmetic comparators~\cite{sanchez2025automatic} as described by the respective boolean formula. Integer representation is selected for simplicity, even so any type of construct following label encoding could be modeled using these same variables. An initial test is conducted for a formula that contains a single construct (integer values) and is in DNF:
\begin{equation}\label{DNF}
\resizebox{\linewidth}{!}{$
\begin{gathered}
\mathcal{F}(x) = \Big[(x>12)\wedge(x<17)\Big] \vee (x=38) \vee \Big[(x>39)\wedge(x<43)\Big] \vee (x\geq56). \\ 
\pmod{64}
\end{gathered}
$}
\end{equation}

\begin{figure}[!hb]
\centerline{\includegraphics[width=0.49\textwidth]{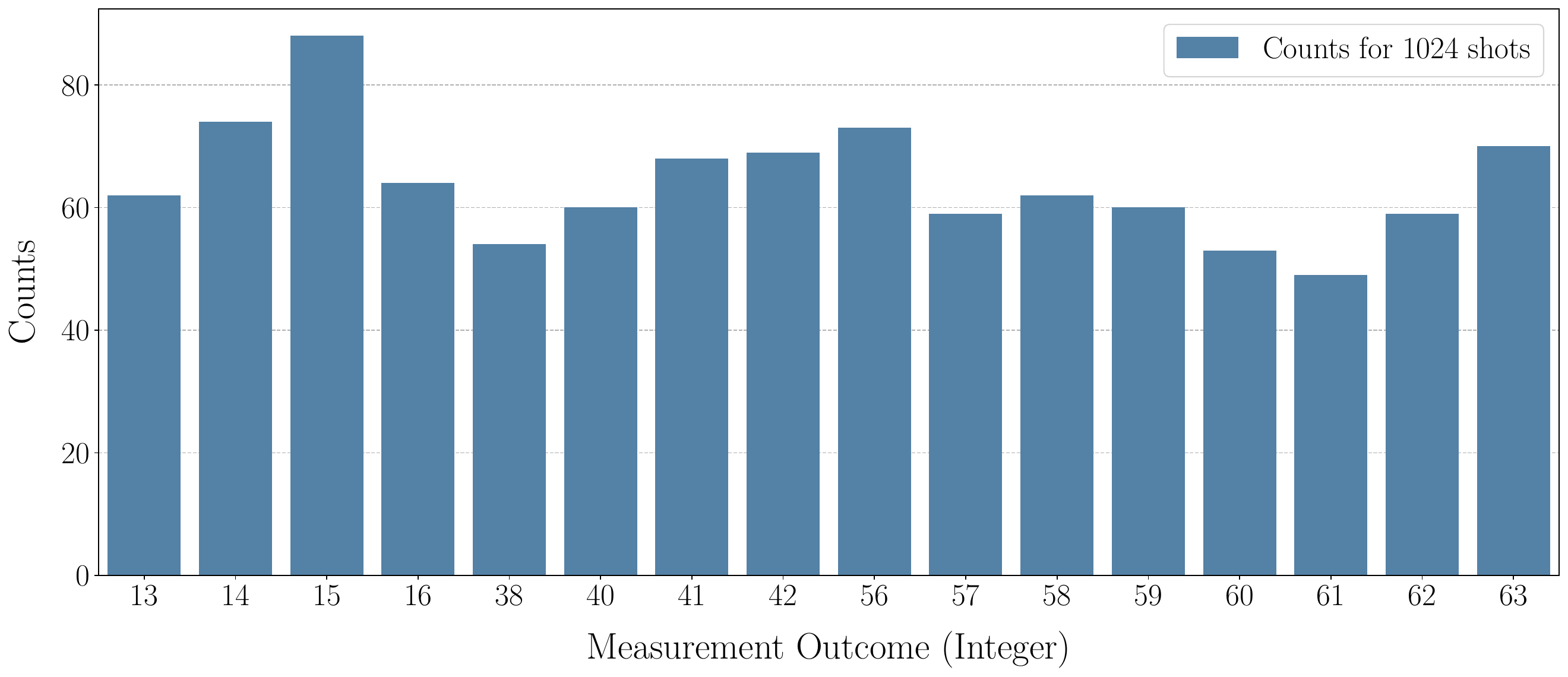}}
\caption{Histogram representing the results obtained after evaluating Algorithm~\ref{alg:quantum_smt_ghz_partitioned} for the case of Eq.~\eqref{DNF} after 1024 \textit{shots}.}\label{SMT:one}
\end{figure}

\begin{figure}[!hb]
    \centering
    \resizebox{\linewidth}{!}{%
    \begin{quantikz}
        \lstick{$\overline{q}_1$} & \qwbundle{7} & \gate[wires=4]{SYNC_{GHZ}} & \gate{\substack{\text{Oracle} \\ x \in [13,17)}} & \gate[wires=4]{Ph_{\vee,\pi}} & \gate{\substack{\text{Oracle} \\ x \in [13,17)}} & \gate[wires=4]{SYNC_{GHZ}^{\dagger}} & \gate{REFL_{\Pi_0}} & \gate[wires=4]{SYNC_{GHZ}} & \meter{} \cw \\
        \lstick{$\overline{q}_2$} & \qwbundle{7} & \ghost{} & \gate{\substack{\text{Oracle} \\ x = 38}} & \ghost{} & \gate{\substack{\text{Oracle} \\ x = 38}} & \ghost{} & & & \\
        \lstick{$\overline{q}_3$} & \qwbundle{7} & \ghost{} & \gate{\substack{\text{Oracle} \\ x \in [40,43)}} & \ghost{} & \gate{\substack{\text{Oracle} \\ x \in [40,43)}} & \ghost{} & & & \\
        \lstick{$\overline{q}_4$} & \qwbundle{7} & \ghost{} & \gate{\substack{\text{Oracle} \\ x \geq 56}} & \ghost{} & \gate{\substack{\text{Oracle} \\ x \geq 56}} & \ghost{} & & &
    \end{quantikz}%
    }
    \caption{Circuit derived from the high-level syntax of Algorithm~\ref{alg:quantum_smt_ghz_partitioned} for the case of Eq.~\eqref{DNF}.}
    \label{circuit}
\end{figure}
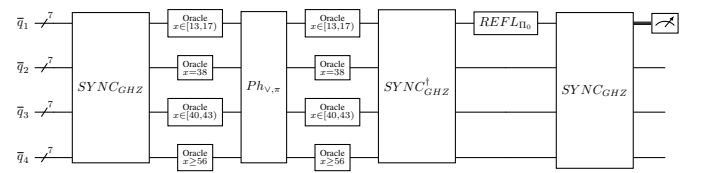

\noindent By performing a Grover's search following the scheme from above, the values that solve the formula, as shown in Fig.~\ref{SMT:one}, are obtained. It is direct to confirm that all values measured correspond to valid solution states for the formula (exact number of Grover's iterations by design). Its associated circuit is illustrated in Fig.~\ref{circuit}.

\begin{figure}[t]
\centerline{\includegraphics[width=0.49\textwidth]{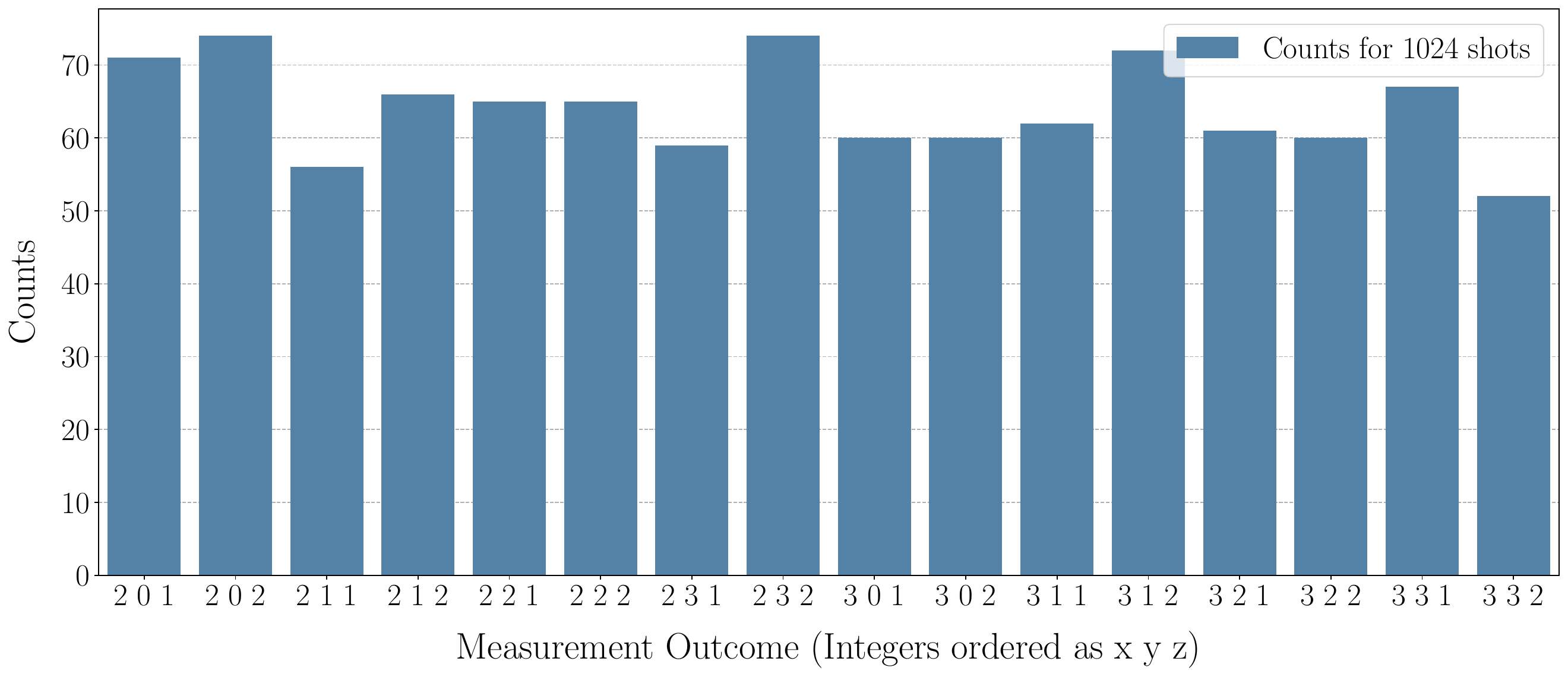}}
\caption{Histogram representing the results obtained after evaluating Algorithm~\ref{alg:quantum_smt_ghz_partitioned} for the case of Eq.~\eqref{CNF} after 1024 \textit{shots}.}\label{SMT:several}
\end{figure}

On the other hand, a second test can be performed on a formula where more than one construct is involved. For instance, formula in CNF:
\begin{equation}\label{CNF}
    \mathcal{F}(x, y, z) = (x \geq 2) \land (y \geq 0) \land (z \geq 1) \land (z<3),
\end{equation}

\noindent with $x$, $y$, and $z$, all evaluated modulo 4. This formula denotes an instance where several constructs are combined in an structured manner. Fig.~\ref{SMT:several} plots results obtained following the program design for the proposed formula.

\section{Discussion}\label{RW}
In this work, we have defined transformations acting on quantum registers using the algebraic formalism introduced by Ying et al.~\cite{ying2026laws}. In comparison, \cite{ying2024verification} uses this same formalism to generalize lower-level abstractions that act on individual qubits rather than registers. Due to its verification properties, the selected formalism also serves as an appropriate tool for transformation formalization as it enables a future framework where routines are verifiable. As partly demonstrated in Section~\ref{abs:compose}, it also provides optimization at a high-level for certain sequences. Hence, obtaining a sequence composed of a simpler structure of operations as a preliminary step prior to translation into the circuit paradigm. However, at this point, this optimization is performed manually as there is no automatization for this process yet. Nonetheless, foundations of a language specification that allows for this direct translation have already been presented~\cite{zhang2025quantum}. If achieved, then it would be possible to find global restructurings that might otherwise not be feasible using circuit gate-based compilers~\cite{xu2022quartz,xu2023synthesizing}.

Essentially, adopting this approach has allowed us to generalize higher-level abstractions that enable the application of composition operations. Programming languages such as \textit{Silq}~\cite{bichsel2020silq}, \textit{Qrisp}~\cite{seidel2024qrisp,bock2025designing} or \textit{Qmod}~\cite{vax2025qmod} are especially promising due to their higher-level of data representation but lack a proper framework for handling them in an structured manner. That is where this contribution takes place, at a stage where composition can be achieved among registers, bringing quantum programming one step closer to high-level structured programming. Notwithstanding, this framework serves principally as a method for quantum program design and, as of now, it is dependent on tracking features of quantum mechanics such as entanglement. As stated by Ali and Yue~\cite{ali2023need}, this poses some limitations. Whereas resources like uncomputation are already actively handle by \textit{Silq} or \textit{Qrisp}, or entanglement tracking in the case of \textit{Twist}~\cite{yuan2022twist}, no actual quantum programming language, to the best of our knowledge, allows the programmer to remain unaware from all these simultaneously. A suitable quantum program design, and subsequent quantum programming language, should handle them in a way where the programmer does not need to resort to designing a quantum program taking them into account. 

\section{Conclusions and Future Work}\label{conclusions}
 
In this work, a conceptual framework where quantum registers act as operands, instead of individual qubits, and operate as single, undivided entities that are treated as computational constructs is introduced. Furthermore, it has succeeded in presenting, through an algebraic formalism, a series of high-level syntax along with their corresponding low-level semantics. The formalization of such transformations is fundamental to the abstraction increase and simplification of quantum program design. This lays the groundwork for the construction of a more suitable quantum programming paradigm in which the programmer works at a higher level, instead of being aware of the lower-level implementation. In essence, this work contributes to establishing, at its early stages, some methodological principles towards realizing a high-level quantum structured programming.

Additionally, this conceptual framework has enabled the design of a quantum program that sits at an abstraction level where no quantum gates are addressed and relies on previously defined composition operations and data encoding operations that are presented as oracles. Future work will set new transformations that contribute to extending the framework in order to design programs for a wider range of problems. In particular, formalization of a method akin to quantum singular value transformation may allow for the integration of polynomial transformations, which serve as the backbone of many quantum computing areas, into an structured setting.

Ultimately, future implementations would benefit from a more appropriate programming language than that used for this work's validation. Migration towards higher-level programming languages like \textit{Silq}, \textit{Qrisp} or \textit{Qmod} will be considered. Although, a programming language that treats this transformations as primitives would be preferred, taking into account that it is paramount that an adequate programming language relieves the programmer from as many quantum characteristics (e.g., entanglement, uncomputation) as possible to further facilitate the construction of quantum programs.

\section*{Disclosure}
The Zenodo repository cited in Section~\ref{SMT} was created with the help of Gemini 3 AI to assist with the structural organization and management of the documentation. The core idea, transformations devising and formalization, quantum SMT program design, and paper drafting, originate exclusively from our (authors) own work.

\balance
\bibliographystyle{ieeetr}
\bibliography{mybibliography}

@article{feynman1982simulating,
  title={Simulating physics with computers},
  author={Feynman, Richard P.},
  journal={International Journal of Theoretical Physics},
  volume={21},
  number={6--7},
  pages={467--488},
  year={1982},
  publisher={Springer},
  doi={10.1007/BF02650179}
}

@article{deutsch1985quantum,
  title={{Quantum theory, the Church--Turing principle and the universal quantum computer}},
  author={Deutsch, David},
  journal={Proceedings of the Royal Society of London. A.},
  volume={400},
  number={1818},
  pages={97--117},
  year={1985},
  publisher={The Royal Society London},
  doi={10.1098/rspa.1985.0070}
}

@article{arute2019quantum,
  title={Quantum supremacy using a programmable superconducting processor},
  author={Arute, Frank and others},
  journal={Nature},
  volume={574},
  number={7779},
  pages={505--510},
  year={2019},
  publisher={Nature Publishing Group UK London},
  doi={10.1038/s41586-019-1666-5}
}

@article{shor1999polynomial,
  author = {Shor, Peter W.},
  title = {Polynomial-Time Algorithms for Prime Factorization and Discrete Logarithms on a Quantum Computer},
  journal = {SIAM Journal on Computing},
  volume = {26},
  number = {5},
  pages = {1484-1509},
  year = {1997},
  doi = {10.1137/S0097539795293172}
}

@inproceedings{grover1996search,
author = {Grover, Lov K.},
title = {A fast quantum mechanical algorithm for database search},
year = {1996},
isbn = {0897917855},
doi = {10.1145/237814.237866},
booktitle = {Proceedings of the Twenty-Eighth Annual ACM Symposium on Theory of Computing},
pages = {212–219}
}

@article{de2021materials,
  title={Materials challenges and opportunities for quantum computing hardware},
  author={De Leon, Nathalie P and others},
  journal={Science},
  volume={372},
  number={6539},
  year={2021},
  publisher={American Association for the Advancement of Science},
  doi={10.1126/science.abb2823}
}

@article{oukaira2025qhd,
  author={Oukaira, Aziz},
  journal={IEEE Access}, 
  title={Quantum Hardware Devices (QHDs): Opportunities and Challenges}, 
  year={2025},
  volume={13},
  number={},
  pages={98229-98241},
  doi={10.1109/ACCESS.2025.3576216}}

@article{dalzell2023quantum,
  title={Quantum algorithms: A survey of applications and end-to-end complexities},
  author={Dalzell, Alexander M and others},
  journal={arXiv preprint arXiv:2310.03011},
  year={2023},
  doi={10.48550/arXiv.2310.03011}
}

@article{murillo2025qse,
author = {Murillo, Juan Manuel and others},
title = {Quantum Software Engineering: Roadmap and Challenges Ahead},
year = {2025},
volume = {34},
number = {5},
issn = {1049-331X},
doi = {10.1145/3712002},
journal = {ACM Transactions on Software Engineering and Methodology},
articleno = {154}
}

@article{abughanem2024ibm,
  title={{IBM quantum computers: evolution, performance, and future directions}},
  author={AbuGhanem, Muhammad},
  journal={arXiv preprint arXiv:2410.00916},
  year={2024},
  doi={10.48550/arXiv.2410.00916}
}

@article{bergholm2022pennyLane,
  title   = {{PennyLane: Automatic differentiation of hybrid quantum-classical computations}},
  author  = {Bergholm, Ville and others},
  year    = {2022},
  journal = {arXiv preprint arXiv:1811.04968},
  doi     = {10.48550/arXiv.1811.04968}
}

@article{qiskit2024,
  title={Quantum computing with {Q}iskit},
  author={Javadi-Abhari, Ali and others},
  year={2024},
  journal={arXiv preprint arXiv:2405.08810},
  doi={10.48550/arXiv.2405.08810},
  eprint={2405.08810},
  archivePrefix={arXiv},
  primaryClass={quant-ph}
}

@article{cirq2025,
  author = {{Cirq Developers}},
  title = {Cirq},
  year = {2025},
  journal = {Zenodo},
  doi = {10.5281/zenodo.4062499},
  url = {https://doi.org/10.5281/zenodo.4062499}
}

@inproceedings{di2025art,
  title={The Art of Abstraction in Quantum Software},
  author={Di Matteo, Olivia},
  booktitle={2025 IEEE/ACM International Workshop on Quantum Software Engineering (Q-SE)},
  pages={25--26},
  year={2025},
  organization={IEEE},
  doi={10.1109/Q-SE66736.2025.00010}
}

@article{liskov1974programming,
  title={Programming with abstract data types},
  author={Liskov, Barbara and Zilles, Stephen},
  journal={ACM SIGPLAN Notices},
  volume={9},
  number={4},
  pages={50--59},
  year={1974},
  doi = {10.1145/942572.807045}
}

@article{clarke1994model,
  title={Model checking and abstraction},
  author={Clarke, Edmund M and Grumberg, Orna and Long, David E},
  journal={ACM Transactions on Programming Languages and Systems (TOPLAS)},
  volume={16},
  number={5},
  pages={1512--1542},
  year={1994},
  publisher={ACM New York, NY, USA},
  doi={10.1145/186025.186051}
}

@book{baldwin2000design,
  title={Design rules, volume 1: The power of modularity},
  author={Baldwin, Carliss Y and Clark, Kim B},
  year={2000},
  publisher={MIT Press},
  doi={10.7551/mitpress/2366.001.0001}
}

@article{shaw2025revisiting,
  title={Revisiting Abstractions for Software Architecture and Tools to Support Them},
  author={Shaw, Mary and Klein, Daniel V and Ross, Theodore L},
  journal={IEEE Transactions on Software Engineering},
  year={2025},
  volume={51},
  number={3},
  pages={768-773},
  publisher={IEEE},
  doi={10.1109/TSE.2025.3533549}
}

@article{shih2021increasing,
  title={Increasing the Level of Abstraction as a Strategy for Accelerating the Adoption of Complex Technologies},
  author={Shih, Willy C},
  journal={Strategy Science},
  volume={6},
  number={1},
  pages={54--61},
  year={2021},
  publisher={INFORMS},
  doi={10.1287/stsc.2020.0116}
}

@inproceedings{di2024abstraction,
  title={An Abstraction Hierarchy Toward Productive Quantum Programming},
  author={Di Matteo, Olivia and Núñez-Corrales, Santiago and Stechły, Micha{\l} and Reinhardt, Steven P and Mattson, Tim},
  booktitle={2024 IEEE International Conference on Quantum Computing and Engineering (QCE)},
  pages={979--989},
  year={2024},
  organization={IEEE},
  doi={10.48550/arXiv.2405.13918}
}

@inproceedings{ali2023need,
  title={On the need of quantum-oriented paradigm},
  author={Ali, Shaukat and Yue, Tao},
  booktitle={Proceedings of the 2nd International Workshop on Quantum Programming for Software Engineering},
  pages={17--20},
  year={2023},
  doi={/10.1145/3617570.3617868}
}

@article{heim2020quantum,
  title={Quantum programming languages},
  author={Heim, Bettina and others},
  journal={Nature Reviews Physics},
  volume={2},
  pages={709--722},
  year={2020},
  publisher={Nature Publishing Group UK London},
  doi={10.1038/s42254-020-00245-7}
}

@inproceedings{chiyao1993,
  author={Chi-Chih Yao, A.},
  booktitle={Proceedings of 1993 IEEE 34th Annual Foundations of Computer Science}, 
  title={Quantum circuit complexity}, 
  year={1993},
  volume={},
  number={},
  pages={352-361},
  doi={10.1109/SFCS.1993.366852}
}

@article{deutsch1989quantum,
  title={Quantum computational networks},
  author={Deutsch, David Elieser},
  journal={Proceedings of the Royal Society of London. A.},
  volume={425},
  number={1868},
  pages={73--90},
  year={1989},
  publisher={The Royal Society London},
  doi={10.1098/rspa.1989.0099}
}

@article{chakraborty2011quect,
  title={{QuECT}: A New Quantum Programming Paradigm},
  author={Chakraborty, Arnab},
  journal={arXiv preprint arXiv:1104.0497},
  year={2011},
  doi={10.48550/arXiv.1104.0497}
}

@article{krantz2019quantum,
  title={A quantum engineer's guide to superconducting qubits},
  author={Krantz, Philip and Kjaergaard, Morten and Yan, Fei and Orlando, Terry P and Gustavsson, Simon and Oliver, William D},
  journal={Applied Physics Reviews},
  volume={6},
  number={2},
  pages={021318},
  year={2019},
  publisher={AIP Publishing},
  doi={10.1063/1.5089550}
}

@article{cleve1998quantum,
  title={Quantum algorithms revisited},
  author={Cleve, Richard and Ekert, Artur and Macchiavello, Chiara and Mosca, Michele},
  journal={Proceedings of the Royal Society of London. A.},
  volume={454},
  number={1969},
  pages={339--354},
  year={1998},
  publisher={The Royal Society},
  doi={10.1098/rspa.1998.0164}
}

@article{ying2026laws,
  title={Laws of Quantum Programming},
  author={Ying, Mingsheng and Zhou, Li and Barthe, Gilles},
  journal={ACM Transactions on Software Engineering and Methodology},
  volume = {35},
  number = {7},
  year={2026},
  articleno = {191},
  numpages = {37},
  publisher={ACM New York, NY},
  doi={10.1145/3765903}
}

@article{hoare1987laws,
  title = {Laws of programming},
  author = {Hoare, C.A.R. and others},
  year = {1987},
  issn = {00010782},
  journal = {Communications of the ACM},
  number = {8},
  pages = {672-686},
  volume = {30},
  doi = {10.1145/27651.27653},
}

@inproceedings{xu2022quartz,
  title={Quartz: superoptimization of Quantum circuits},
  author={Xu, Mingkuan and others},
  booktitle={Proceedings of the 43rd ACM SIGPLAN International Conference on Programming Language Design and Implementation},
  pages={625--640},
  year={2022},
  doi={10.1145/3519939.3523433}
}

@article{xu2023synthesizing,
  title={Synthesizing quantum-circuit optimizers},
  author={Xu, Amanda and Molavi, Abtin and Pick, Lauren and Tannu, Swamit and Albarghouthi, Aws},
  journal={Proceedings of the ACM on Programming Languages},
  volume={7},
  number={PLDI},
  pages={835--859},
  year={2023},
  publisher={ACM New York, NY, USA},
  doi={10.1145/3591254}
}

@inproceedings{lin2024parallel,
  title={A Parallel and Distributed Quantum SAT Solver Based on Entanglement and Teleportation},
  author={Lin, Shang-Wei and Wang, Tzu-Fan and Chen, Yean-Ru and Hou, Zhe and San{\'a}n, David and Teo, Yon Shin},
  booktitle={International Conference on Tools and Algorithms for the Construction and Analysis of Systems},
  pages={363--382},
  year={2024},
  organization={Springer},
  doi={10.1007/978-3-031-57249-4_18}
}

@inproceedings{bichsel2020silq,
  title={Silq: A high-level quantum language with safe uncomputation and intuitive semantics},
  author={Bichsel, Benjamin and Baader, Maximilian and Gehr, Timon and Vechev, Martin},
  booktitle={Proceedings of the 41st ACM SIGPLAN Conference on Programming Language Design and Implementation},
  pages={286--300},
  year={2020},
  doi={10.1145/3385412.3386007}
}

@article{yuan2022twist,
  title={Twist: Sound reasoning for purity and entanglement in quantum programs},
  author={Yuan, Charles and McNally, Christopher and Carbin, Michael},
  journal={Proceedings of the ACM on Programming Languages},
  volume={6},
  number={POPL},
  pages={1--32},
  year={2022},
  publisher={ACM New York, NY, USA},
  doi={10.1145/3498691}
}

@article{ying2024verification,
  title={Verification of Recursively Defined Quantum Circuits},
  author={Ying, Mingsheng and Zhang, Zhicheng},
  journal={arXiv preprint arXiv:2404.05934},
  year={2024},
  doi={10.48550/arXiv.2404.05934}
}

@article{zhang2025quantum,
  title={Quantum Register Machine: Efficient Implementation of Quantum Recursive Programs},
  author={Zhang, Zhicheng and Ying, Mingsheng},
  journal={Proceedings of the ACM on Programming Languages},
  volume={9},
  number={PLDI},
  pages={822--847},
  year={2025},
  publisher={ACM New York, NY, USA},
  doi={10.1145/3729283}
}

@book{wilkes1958preparation,
  title={The preparation of programs for an electronic digital computer},
  author={Wilkes, Maurice V and Wheeler, David J and Gill, Stanley},
  year={1957},
  publisher={Addison-Wesley Press},
  edition={2nd}
}

@article{seidel2024qrisp,
  title={Qrisp: A Framework for Compilable High-Level Programming of Gate-Based Quantum Computers},
  author={Seidel, Raphael and others},
  journal={arXiv preprint arXiv:2406.14792},
  year={2024},
  doi={10.48550/arXiv.2406.14792}
}

@conference{bock2025designing,
  author={Sebastian Bock and Raphael Seidel and Matic Petrič and Nikolay Tcholtchev and Andreas Hoffmann and Niklas Porges},
  title={Designing a Meta-Model for the Eclipse Qrisp eDSL for High-Level Quantum Programming},
  booktitle={Proceedings of the 13th International Conference on Model-Based Software and Systems Engineering - MODELSWARD},
  year={2025},
  volume={1},
  pages={27-39},
  publisher={SciTePress},
  doi={10.5220/0013121000003896}
}

@article{sanchez2025automatic,
  title={Automatic generation of efficient oracles: The less-than case},
  author={Sanchez-Rivero, Javier and Talav{\'a}n, Daniel and Garcia-Alonso, Jose and Ruiz-Cort{\'e}s, Antonio and Murillo, Juan Manuel},
  journal={Journal of Systems and Software},
  volume={219},
  pages={112203},
  year={2025},
  publisher={Elsevier},
  doi={10.1016/j.jss.2024.112203}
}

@article{black2013object,
  title={Object-oriented programming: Some history, and challenges for the next fifty years},
  author={Black, Andrew P},
  journal={Information and Computation},
  volume={231},
  pages={3--20},
  year={2013},
  publisher={Elsevier},
  doi={10.48550/arXiv.1303.0427}
}

@article{bennett1973reversibility,
  author={Bennett, C. H.},
  journal={IBM Journal of Research and Development}, 
  title={Logical Reversibility of Computation}, 
  year={1973},
  volume={17},
  number={6},
  pages={525-532},
  doi={10.1147/rd.176.0525}
}

@article{brassard2000quantum,
  title={Quantum amplitude amplification and estimation},
  author={Brassard, Gilles and Hoyer, Peter and Mosca, Michele and Tapp, Alain},
  journal={arXiv preprint quant-ph/0005055},
  year={2000},
  doi={10.48550/arXiv.quant-ph/0005055}
}

@inproceedings{szegedy2004quantum,
  author={Szegedy, M.},
  booktitle={45th Annual IEEE Symposium on Foundations of Computer Science}, 
  title={Quantum speed-up of Markov chain based algorithms}, 
  year={2004},
  volume={},
  number={},
  pages={32-41},
  doi={10.1109/FOCS.2004.53}
}

@article{biamonte2017quantum,
  title={Quantum machine learning},
  author={Biamonte, Jacob and Wittek, Peter and Pancotti, Nicola and Rebentrost, Patrick and Wiebe, Nathan and Lloyd, Seth},
  journal={Nature},
  volume={549},
  pages={195--202},
  year={2017},
  publisher={Nature Publishing Group UK London},
  doi={10.1038/nature23474}
}

@article{rebentrost2014quantum,
  title = {Quantum Support Vector Machine for Big Data Classification},
  author = {Rebentrost, Patrick and Mohseni, Masoud and Lloyd, Seth},
  journal = {Physical Review Letters},
  volume = {113},
  issue = {13},
  pages = {130503},
  numpages = {5},
  year = {2014},
  publisher = {American Physical Society},
  doi = {10.1103/PhysRevLett.113.130503}
}

@inproceedings{gilyen2019quantum,
  title={Quantum singular value transformation and beyond: exponential improvements for quantum matrix arithmetics},
  author={Gily{\'e}n, Andr{\'a}s and Su, Yuan and Low, Guang Hao and Wiebe, Nathan},
  booktitle={Proceedings of the 51st Annual ACM SIGACT Symposium on Theory of Computing},
  pages={193--204},
  year={2019},
  doi={10.1145/3313276.3316366}
}

@article{martyn2021grand,
  title = {Grand Unification of Quantum Algorithms},
  author = {Martyn, John M. and Rossi, Zane M. and Tan, Andrew K. and Chuang, Isaac L.},
  journal = {PRX Quantum},
  volume = {2},
  number = {4},
  pages = {040203},
  numpages = {40},
  year = {2021},
  publisher = {American Physical Society},
  doi = {10.1103/PRXQuantum.2.040203}
}

@article{low2019hamiltonian,
  title={Hamiltonian simulation by qubitization},
  author={Low, Guang Hao and Chuang, Isaac L},
  journal={Quantum},
  volume={3},
  pages={163},
  year={2019},
  publisher={Verein zur F{\"o}rderung des Open Access Publizierens in den Quantenwissenschaften},
  doi={10.22331/q-2019-07-12-163}
}

@incollection{greenberger1989going,
  title={{Going beyond Bell’s theorem}},
  author={Greenberger, Daniel M and Horne, Michael A and Zeilinger, Anton},
  booktitle={Kafatos, M. (eds) Bell’s Theorem, Quantum Theory and Conceptions of the Universe. Fundamental Theories of Physics},
  volume={37},
  pages={69--72},
  year={1989},
  publisher={Springer},
  doi={10.48550/arXiv.0712.0921}
}

@article{wootters1982single,
  title={A single quantum cannot be cloned},
  author={Wootters, William K and Zurek, Wojciech H},
  journal={Nature},
  volume={299},
  pages={802--803},
  year={1982},
  publisher={Nature Publishing Group UK London},
  doi={10.1038/299802a0}
}

@article{rosa2025optimizing,
  title={Optimizing gate decomposition for high-level quantum programming},
  author={Rosa, Evandro CR and Duzzioni, Eduardo I and De Santiago, Rafael},
  journal={Quantum},
  volume={9},
  pages={1659},
  year={2025},
  publisher={Verein zur F{\"o}rderung des Open Access Publizierens in den Quantenwissenschaften},
  doi={10.22331/q-2025-03-12-1659}
}

@inproceedings{balauca2022efficient,
  title={Efficient constructions for simulating multi controlled quantum gates},
  author={Balauca, Stefan and Arusoaie, Andreea},
  booktitle={International Conference on Computational Science},
  pages={179--194},
  year={2022},
  organization={Springer},
  doi={10.1007/978-3-031-08760-8_16}
}

@article{nie2024quantum,
  title={{Quantum circuit for multi-qubit Toffoli gate with optimal resource}},
  author={Nie, Junhong and Zi, Wei and Sun, Xiaoming},
  journal={arXiv preprint arXiv:2402.05053},
  year={2024},
  doi={10.48550/arXiv.2402.05053}
}

@article{barenco1995elementary,
  title={Elementary gates for quantum computation},
  author={Barenco, Adriano and others},
  journal={Physical Review A},
  volume={52},
  pages={3457},
  year={1995},
  publisher={APS},
  doi={10.1103/PhysRevA.52.3457}
}

@article{grover2005fixed,
  title={Fixed-point quantum search},
  author={Grover, Lov K},
  journal={Physical Review Letters},
  volume={95},
  number={15},
  pages={150501},
  year={2005},
  publisher={APS},
  doi={10.1103/PhysRevLett.95.150501}
}

@article{yoder2014fixed,
  title={Fixed-point quantum search with an optimal number of queries},
  author={Yoder, Theodore J and Low, Guang Hao and Chuang, Isaac L},
  journal={Physical Review Letters},
  volume={113},
  number={21},
  pages={210501},
  year={2014},
  publisher={APS},
  doi={10.1103/PhysRevLett.113.210501}
}

@book{bradley2007calculus,
  title={The calculus of computation: decision procedures with applications to verification},
  author={Bradley, Aaron R and Manna, Zohar},
  year={2007},
  publisher={Springer},
  doi={10.1007/978-3-540-74113-8}
}

@article{de2011satisfiability,
  title={Satisfiability modulo theories: introduction and applications},
  author={De Moura, Leonardo and Bj{\o}rner, Nikolaj},
  journal={Communications of the ACM},
  volume={54},
  number={9},
  pages={69--77},
  year={2011},
  publisher={ACM New York, NY, USA},
  doi={10.1145/1995376.1995394}
}

@incollection{barrett2018satisfiability,
  title={Satisfiability modulo theories},
  author={Barrett, Clark and Tinelli, Cesare},
  booktitle={Handbook of Model Checking},
  pages={305--343},
  year={2018},
  publisher={Springer},
  doi={10.1007/978-3-319-10575-8_11}
}

@article{jiang2026correction,
  title={Correction to: Advancements in superconducting quantum computing},
  author={Jiang, Yao-Yao and others},
  journal={National Science Review},
  volume={13},
  number={2},
  pages={nwaf582},
  year={2026},
  publisher={Oxford University Press},
  doi={10.1093/nsr/nwaf582}
}

@article{zhao2020quantum,
  title={Quantum software engineering: Landscapes and horizons},
  author={Zhao, Jianjun},
  journal={arXiv preprint arXiv:2007.07047},
  year={2020},
  doi={10.48550/arXiv.2007.07047}
}

@article{moore2001parallel,
  title={Parallel quantum computation and quantum codes},
  author={Moore, Cristopher and Nilsson, Martin},
  journal={SIAM Journal on Computing},
  volume={31},
  number={3},
  pages={799--815},
  year={2001},
  publisher={SIAM},
  doi={10.1137/S0097539799355053}
}

@techreport{abhari2012scaffold,
  title={Scaffold: Quantum programming language},
  author={Abhari, Ali J and others},
  institution={Princeton University},
  year={2012}
}

@inproceedings{svore2018q,
  title={Q\# enabling scalable quantum computing and development with a high-level dsl},
  author={Svore, Krysta and others},
  booktitle={Proceedings of the Real World Domain Specific Languages Workshop 2018},
  pages={1--10},
  number={7},
  year={2018},
  doi={10.1145/3183895.3183901}
}

@article{green2013quipper,
  title={Quipper: a scalable quantum programming language},
  author={Green, Alexander S and Lumsdaine, Peter LeFanu and Ross, Neil J and Selinger, Peter and Valiron, Beno{\^\i}t},
  volume = {48},
  number = {6},
  journal = {ACM SIGPLAN Notices},
  pages = {333–342},
  numpages = {10},
  publisher = {Association for Computing Machinery},
  year={2013},
  doi={10.1145/2491956.2462177}
}

@book{ying2024foundations,
  title={Foundations of quantum programming},
  author={Ying, Mingsheng},
  year={2024},
  edition={2nd},
  publisher={Morgan Kaufmann by Elsevier}
}

@article{vedral1996quantum,
  title={Quantum networks for elementary arithmetic operations},
  author={Vedral, Vlatko and Barenco, Adriano and Ekert, Artur},
  journal={Physical Review A},
  volume={54},
  number={1},
  pages={147},
  year={1996},
  publisher={APS},
  doi={10.1103/PhysRevA.54.147}
}

@article{draper2000addition,
  title={Addition on a quantum computer},
  author={Draper, Thomas G},
  journal={arXiv preprint quant-ph/0008033},
  year={2000},
  doi={10.48550/arXiv.quant-ph/0008033}
}

@article{dervovic2018quantum,
  title={Quantum linear systems algorithms: a primer},
  author={Dervovic, Danial and Herbster, Mark and Mountney, Peter and Severini, Simone and Usher, Na{\"\i}ri and Wossnig, Leonard},
  journal={arXiv preprint arXiv:1802.08227},
  year={2018},
  doi={10.48550/arXiv.1802.08227}
}

@inproceedings{zhao2025abstraction,
  author={Zhao, Jianjun},
  booktitle={2025 40th IEEE/ACM International Conference on Automated Software Engineering (ASE)}, 
  title={When Abstraction Breaks Physics: Rethinking Modular Design in Quantum Software}, 
  year={2025},
  volume={},
  number={},
  pages={3886-3890},
}

@article{chamizozenodo,
  author={Chamizo, David and Garcia-Alonso, Jose and Murillo, Juan M.},
  title={High-level quantum structured programs as quantum registers compositions},
  year=2026,
  journal={{Zenodo}},
  version={1.0.1},
  doi={10.5281/zenodo.19762070},
  note={https://doi.org/10.5281/zenodo.19762069 (Accessed July 2026)},
}

@article{brun_2020, 
  title={Quantum Error Correction}, 
  ISBN={9780197851753}, 
  DOI={10.1093/acrefore/9780190871994.013.35}, 
  journal={Oxford Research Encyclopedia of Physics}, 
  publisher={Oxford University PressNew York, NY}, 
  author={Brun, Todd A},
  year={2020}
}

@article{vax2025qmod,
  title={Qmod: Expressive high-level quantum modeling},
  author={Vax, Matan and others},
  journal={arXiv preprint arXiv:2502.19368},
  year={2025}
}

@article{goldfriend2025design,
  title={Design and synthesis of scalable quantum programs},
  author={Goldfriend, Tomer and others},
  journal={arXiv preprint arXiv:2412.07372},
  year={2025}
}

@article{yan2025quantum,
  title={Quantum circuit synthesis and compilation optimization: Overview and prospects},
  author={Yan, Ge and others},
  journal={arXiv preprint arXiv:2407.00736},
  year={2025}
}

@article{mckay2017efficient,
    title = {{Efficient Z gates for quantum computing}},
    year = {2017},
    journal = {Physical Review A},
    author = {McKay, David C. and Wood, Christopher J. and Sheldon, Sarah and Chow, Jerry M. and Gambetta, Jay M.},
    volume = {96},
    publisher = {American Physical Society},
    url = {https://journals.aps.org/pra/abstract/10.1103/PhysRevA.96.022330},
    doi = {10.1103/PhysRevA.96.022330},
    issn = {24699934},
    arxivId = {1612.00858},
    pages={022330}
}

@article{da2022linear,
  title={Linear-depth quantum circuits for multiqubit controlled gates},
  author={Da Silva, Adenilton J and Park, Daniel K},
  journal={Physical Review A},
  volume={106},
  pages={042602},
  year={2022},
  publisher={APS}
}

\end{document}